\magnification=1200
\input amstex
\input epsf  
\documentstyle{amsppt}
\pagewidth{32pc}
\pageheight{45.5pc}
\overfullrule=0pt
\nologo
\UseAMSsymbols

\input miniltx
 \expandafter\def\expandafter\+\expandafter{\+}
 \input url.sty

\centerline{\bf  FOCK SPACE: A BRIDGE BETWEEN}
\centerline{\bf  FREDHOLM INDEX AND THE QUANTUM HALL EFFECT}

\bigskip

\centerline{\bf  Guo Chuan Thiang and Jingbo Xia}

{\input amsppt.sty
\topmatter
\thanks
\endthanks
\endtopmatter
}

\noindent
{\bf Abstract.}   We compute the quantized Hall conductance at various Landau levels by using the classic trace.
The computations reduce to the single elementary one for the lowest Landau level.
By using the theories of Helton-Howe-Carey-Pincus, and Toeplitz operators on the classic Fock space and higher Fock spaces,  
the Hall conductance is naturally identified with a Fredholm index.   This brings new mathematical 
insights to the extraordinary precision of quantization observed in quantum Hall measurements.

\bigskip
\centerline{\bf 1.  Introduction}
\medskip
Consider an electron confined to a plane with a perpendicular magnetic field  ${\bold B}$  of uniform strength.  
Under the right choice of orientation, identify the plane with  ${\bold C}$,  the complex plane.  
In the symmetric gauge,  we have the free Hamiltonian   
$$
H_b =  \bigg({1\over i}{\partial \over \partial x} + {b\over 2}y\bigg)^2  +  \bigg({1\over i}{\partial \over \partial y} - {b\over 2}x\bigg)^2
$$
\noindent
representing this system,  where  $b = e|{\bold B}|/\hbar c > 0$.  
When the Fermi energy  $E$  is in a spectral gap of  $H_b$,  
we have the Fermi projection  $P_{\leq E} = \chi _{(-\infty ,E]}(H_b)$.  Let  $f_1$  and  $f_2$  be {\it switch functions} in the  
$x$  and  $y$  directions respectively  (see Definition 3.5 below).    It is known  that the expression 
$$
\sigma _{\text{Hall}}(P_{\leq E})  =  -i\text{tr}(P_{\leq E}[[f_1,P_{\leq E}],[f_2,P_{\leq E}]])
\tag 1.1
$$
\noindent
is the Kubo formula for the Hall conductance of  $P_{\leq E}$, provided that the trace on the right-hand side makes sense.  See [7,1].
\medskip
The spectrum of  $H_b$  is well known.  In fact, an easy diagonalization of  $H_b$  shows that its spectrum consists of the 
eigenvalues  $\{(2\ell + 1)b : \ell  = 0, 1, 2, \dots \}$,  called Landau levels, each with infinite degeneracy.    
In the complex coordinates, it is easy to verify that
$$
H_b - b  =  4(-\partial +(b/4)\bar z)(\bar \partial + (b/4)z),
$$
\noindent
where  $\partial = \partial /\partial z$  and  $\bar \partial = \partial /\partial \bar z$.
From this we see that the eigenspace corresponding to the lowest Landau level (LLL) is the closure in  $L^2({\bold C})$  of 
$$
\text{span}\{z^ke^{-(b/4)|z|^2} :  k = 0, 1, 2, \dots \}.  
\tag 1.2
$$
\noindent
One immediately recognizes that this is a unitarily equivalent form of the familiar Fock space.
Thus in the study of the quantum Hall effect,
the wealth of knowledge that has been accumulated in the study of operators on the Fock space can be brought to bear.
\medskip
Define the measure  
$$
d\mu (z)  =  {1\over \pi }e^{-|z|^2}dA(z)
$$
\noindent
on  ${\bold C}$,  where  $dA$  is the standard area measure.  Recall that the Fock space  ${\Cal F}^2$  is the closure of  
the analytic polynomials  ${\bold C}[z]$  in  $L^2({\bold C},d\mu )$.   As usual, we write  $P$  for the orthogonal projection 
from  $L^2({\bold C},d\mu )$  onto  ${\Cal F}^2$.
\medskip
Define  
$$
(U_b\psi )(z)  =  (2/b)^{1/2}\psi ((2/b)^{1/2}z)e^{|z|^2/2},  \quad    \psi  \in L^2({\bold C},dA).
$$
\noindent
Then  $U_b$ is a unitary operator that maps  $L^2({\bold C},dA)$  onto   $L^2({\bold C},d\mu )$.  
Let  ${\Cal L}$  be the eigenspace for  $H_b$  corresponding to the eigenvalue   $b$.  By (1.2),  we have
$$
U_b{\Cal L}  =  {\Cal F}^2.
$$
\noindent
Moreover,
$$
U_bP_{\leq E}U_b^\ast  =  P   \quad  \text{when}  \ \  b < E < 3b.
$$
\noindent
Combining this with (1.1),  we see that when  $b < E < 3b$,
$$
\sigma _{\text{Hall}}(P_{\leq E})  =  -i\text{tr}(P[[M_{\hat f_1},P],[M_{\hat f_2},P]]),
\tag 1.3
$$
\noindent
where  $\hat f_i(z) = f_i((2/b)^{1/2}z)$,  and  $M_{\hat f_i}$  is the operator of multiplication by the function  
$\hat f_i$  on  $L^2({\bold C},d\mu )$,  $i = 1,2$.  This realizes 
realizes the Hall conductance of the lowest Landau level as a trace on the 
Fock space.  We can further rewrite the right-hand side in terms of Toeplitz operators.
\medskip
Recall that for each  $f \in L^\infty ({\bold C})$,  the Toeplitz operator  $T_f$  is defined by the formula
$$
T_fh  =  P(fh),  \quad  h \in {\Cal F}^2.
$$
\noindent
For any  $f, g \in L^\infty ({\bold C})$,  we have
$$
\align
[T_f,T_g]  &=  PM_fPM_gP - PM_gPM_fP   \\
&=  P(M_fP - PM_f)(M_gP - PM_g) - P(M_gP - PM_g)(M_fP - PM_f)   \\
&= P[[M_f,P],[M_g,P]].
\tag 1.4
\endalign
$$
\noindent
Therefore,  renaming  $\hat f_1$, $\hat f_2$  as  $f_1$,  $f_2$,  when  $b < E < 3b$,  (1.3)  becomes
$$
\sigma _{\text{Hall}}(P_{\leq E})  =  -i\text{tr}[T_{f_1},T_{f_2}].
\tag 1.5
$$
\noindent
In an earlier work [11], the first author showed, by an explicit calculation, that
$$
\text{tr}[T_{f_1},T_{f_2}] = \text{tr}(P[[M_{f_1},P],[M_{f_2},P]]) = {1\over 2\pi i},
\tag 1.6
$$
\noindent
independently of the choice of switch functions.
This obtains the value  $-1/2\pi $  for the Hall conductance  $\sigma _{\text{Hall}}(P_{\leq E})$,  which gives a  
mathematical proof to what was a folklore in physics.
\medskip
In the decades since the initial discovery of the quantum Hall effect, repeated experiments have 
seen an extraordinary stability for the observed integer values of  $2\pi \sigma _{\text{Hall}}$.  
This phenomenon calls for an explanation in terms of a mathematical quantity that is stable under small perturbations  
and quantized to integer values.   In this regard, the Fredholm index is an obvious candidate.
\medskip
There were previous attempts along this line, one of the earliest being [1].  In Theorem 6.8 of that paper, 
Avron, Seiler and Simon used their theory of {\it relative index of projections} to show that
$$
2\pi i\text{Tr}[T_{f_1},T_{f_2}]  =  -\text{index}(T_{z/|z|})
\tag 1.7
$$
\noindent
and checked that the right-hand side was  $1$.   
But in [1], the left-hand side of (1.7) was treated as a limit of truncated traces,    
not the classic trace.  (See the comment after [1,(6.6)].)   This leaves room for improvement.  
For example, if the commutator  $[T_{f_1},T_{f_2}]$  is known to be in the trace class, 
we can directly deduce its stability with respect to,  e.g.,  trace class perturbations of  $P$, without using the right-hand side.   
\medskip
Improvements started with [9], in which Ludewig and the first author showed, using general abstract methods,
that the commutator  $[T_{f_1},T_{f_2}]$   is indeed in the trace class.  (Later in Proposition 3.7,
we will verify this trace-class condition more directly.)  Then in [11], the first author did the trace calculation (1.6).
\medskip
In this paper we take the next step.      
Our starting point is (1.5).  Recall that the values of the 
switch functions  $f_1$  and  $f_2$  are contained in  $[0,1]$.  The fact that  $[T_{f_1},T_{f_2}]$  is in the trace class means that 
the theories of Helton-Howe [8] and Carey-Pincus [3,4,10] come into play.  Define  $F = f_1 + if_2$.  
Using a new technique for constructing Fredholm inverses, we will show in Theorem 4.1 that the essential 
spectrum of the Toeplitz operator  $T_F$  is the boundary  $\partial S$  of the square  
$$
S  =  \{x+iy : 0 \leq x  \leq 1 \ \text{and} \ 0 \leq y \leq 1\}.
$$
\noindent
Knowing that the two-dimensional Lebesgue measure of this essential spectrum is  $0$,   
we deduce from the Carey-Pincus theory that the {\it principal function}  $g$  for the almost commuting pair  
$T_{f_1}$,  $T_{f_2}$  is given by the formula
$$
g(x,y) = n\chi _S(x+iy),
$$
\noindent
where
$$
n = \text{index}(T_F - \lambda )  \quad \text{for every}  \ \  \lambda \in S\backslash \partial S.
$$
\noindent
The Carey-Pincus theory also tells us that
$$
\text{tr}[T_{f_1},T_{f_2}]  =  {-1\over 2\pi i}\iint g(x,y)dxdy = {-n\over 2\pi i}.
\tag 1.8
$$
\noindent
Combining this with (1.5),  we find that when  $b < E < 3b$, 
$$
\sigma _{\text{Hall}}(P_{\leq E})  = {1\over 2\pi }\text{index}(T_F - \lambda )  
\quad \text{for every}  \ \  \lambda \in S\backslash \partial S.
\tag 1.9
$$
\noindent
This identifies the Hall conductance with a bona fide 
Fredholm index in a direct way, distinct from the Avron-Seiler-Simon approach leading to (1.7).
Moreover, (1.8) turns (1.6), a trace calculation,  into an index calculation:
$$
\text{index}(T_F - \lambda ) = -1  \quad \text{for every}  \ \  \lambda \in S\backslash \partial S.
$$
\noindent
Consequently,  $g(x,y) = -\chi _S(x+iy)$.   Thus the Carey-Pincus theory gives us the formula
$$
\text{tr}[p(T_{f_1},T_{f_2}),q(T_{f_1},T_{f_2})]  =  {1\over 2\pi i}\iint _S\{p,q\}(x,y)dxdy 
\tag 1.10
$$
\noindent
for  $p, q \in {\bold C}[x,y]$,  where  $\{p,q\}$  is the Poisson bracket of the pair  $p$,  $q$.  In other words, the one 
trace calculation (1.6), plus the determination of the essential spectrum of  $T_F$,  
plus the Carey-Pincus theory, give us all the explicit traces above. 
\medskip
It is worth stressing that even something like  $\text{tr}([T_{f_1},T_{f_2}]T_{f_2})$  would be impossible to calculate 
without the Carey-Pincus theory.   But with the Carey-Pincus theory, not only do we know that  
$\text{tr}([T_{f_1},T_{f_2}]T_{f_2}) = (4\pi i)^{-1}$,  every trace in (1.10) is at our finger tips.  
\medskip
As the main part of our paper,  we will prove the analogues of (1.6), (1.9) and (1.10)  when   $E$  is between {\it arbitrary} 
Landau levels   $(2\ell +1)b$  and  $(2\ell +3)b$,  $\ell \geq 0$.   We emphasize that our aim is to determine, 
for the general case of  $\ell \geq 0$,  the full principal function, which is a lot more than the calculation of 
an individual trace or index.
\medskip
After establishing the general version of  (1.9),  we will take a look at another pair of functions,  inspired by  (1.7).  
We consider  
$$
\varphi _1(z)  =  \text{Re}(z/|z|)   \quad  \text{and}  \quad
\varphi _2(z)  =  \text{Im}(z/|z|),
$$
\noindent
$z \in {\bold C}\backslash \{0\}$,  which are not the kind of switch functions suitable for the Kubo formula,  
but are still worth investigating on mathematical grounds.   Define  $\Phi = \varphi _1 + i\varphi _2$.  That is,  
$$
\Phi (z) = z/|z|
$$  
\noindent
for  $z \in {\bold C}\backslash \{0\}$.  We will show that the Toeplitz operator   $T_\Phi $  on  ${\Cal F}^2$    
is a compact perturbation of the {\it unilateral shift}.  Consequently, the essential spectrum of  $T_\Phi $  equals the 
unit circle  ${\bold T}$.  We will also show that  $[T_{\varphi _1},T_{\varphi _2}]$  is in the trace class,  and that  
$\text{tr}[T_{\varphi _1},T_{\varphi _2}] = (2i)^{-1}$.  This leads to the identity
$$
\text{tr}(P[[M_{\varphi _1},P],[M_{\varphi _2},P]])  =  - {1\over 2i}\text{index}(T_\Phi - \lambda )  \quad \text{for every}  \ \  \lambda \in D,
$$
\noindent
where  $D = \{z \in {\bold C} : |z| < 1\}$.  Furthermore,  we have
$$
\text{tr}[p(T_{\varphi _1},T_{\varphi _2}),q(T_{\varphi _1},T_{\varphi _2})]  
=  {1\over 2\pi i}\iint _D\{p,q\}(x,y)dxdy  
$$
\noindent
for all  $p,q \in {\bold C}[x,y]$. 
\medskip
After that, we will work out a  ``non-Toeplitz" example in Section 10.   In addition to the value of the    
example itself, this goes some way to demonstrate that our techniques are quite general.
\medskip
It should be mentioned that in addition to (1.3),  there are at least two other versions of 
``Kubo formula for Hall conductance"  studied in the literature.  If one uses ``trace-per-unit-area",
$$
{\Cal T}(A)  =  \lim _{\Lambda \rightarrow {\bold C}}{1\over |\Lambda |}\text{tr}(\chi _\Lambda A\chi _\Lambda ),
$$
\noindent
then the quantity  
$$
\sigma _{\text{Hall}}^{(2)}  =  -i{\Cal T}(P[[\hat x,P],[\hat y,P]]),
$$
\noindent
where  $\hat x$  and  $\hat y$  are the coordinate functions in the $x$ and $y$ directions, 
is a second version of Hall conductance.  
\medskip
The quantity  $\sigma _{\text{Hall}}^{(2)}$  fits very nicely in the context of non-commuative geometry [2].  
Thus the study of   $\sigma _{\text{Hall}}^{(2)}$  benefits from the tremendous progress that has been made 
in non-commuative geometry in the last few decades.  
The equality of  $\sigma _{\text{Hall}}$ and  $\sigma _{\text{Hall}}^{(2)}$ 
can be deduced once the trace class membership needed for the former is established.   See [6, Appendix Lemma 8] 
for such arguments in the context of discrete models.  Otherwise, if only the limit of truncated traces is used in defining 
$\sigma _{\text{Hall}}$, then the magnetic translation invariance of  $H_b$, thus of  $P$, needs to be invoked to obtain 
equality with  $\sigma _{\text{Hall}}^{(2)}$, as explained on page 416 of [1].
\medskip
For a third version, consider a Hamiltonian  $H$  that is periodic 
with respect to the lattice  ${\bold Z}^2$  in  ${\bold R}^2$.  Under additional assumptions,  $P$  decomposes 
in the form  $\int _{{\bold T}^2}^\oplus P(k)dm(k)$,  where  ${\bold T}^2$ is the dual Brillouin torus to  ${\bold Z}^2$,  
and each $P(k)$ is a finite-rank projection matrix depending smoothly on $k \in {\bold T}^2$.  
That is,  $P$  is a map from  ${\bold T}^2$  to  $M_{n\times n}$.  
Then a third version of Hall conductance is given by the formula
$$
\sigma _{\text{Hall}}^{(3)}  =  -i\int _{{\bold T}^2}\text{tr}(P[\partial _{k_1}P,\partial _{k_2}P])dm(k),
$$
\noindent
which may be recognized as a Chern number of the vector bundle over ${\bold T}^2$ formed by the ranges of  $P(k)$. 
\medskip
We emphasize that the focus of this paper is on the Hall conductance given by the classic trace, as in (1.1).  
Operator theory is our main tool for analyzing (1.3) and its analogue for higher Landau levels.  
Like the field of non-commutative geometry, tremendous progress has also been made in operator theory on 
reproducing-kernel Hilbert spaces.  This paper benefits from this progress in the sense that our results  
are really applications of this particular form of operator theory.   

\bigskip
\centerline{\bf 2.  Landau levels and eigenspaces}
\medskip
As we already mentioned in the Introduction, the magnetic Hamiltonian
$$
H_b =  \bigg({1\over i}{\partial \over \partial x} + {b\over 2}y\bigg)^2  +  \bigg({1\over i}{\partial \over \partial y} - {b\over 2}x\bigg)^2
$$
\noindent
admits the factorization  
$$
H_b - b  =  4(-\partial +(b/4)\bar z)(\bar \partial + (b/4)z).
$$
\noindent
Let us denote
$$
\quad  \tilde A = 2\bar \partial + (b/2)z   \quad \text{and}  \quad \tilde C  =  -2\partial + (b/2)\bar z.
$$
\noindent
Then
$$
H_b - b  =  \tilde C\tilde A.
$$
\noindent
This pair of operators   $\tilde A$,  $\tilde C$  are called annihilation and creation operators, 
and gives us an explicit diagonalization of  $H_b$.  
Indeed this diagonalization is a standard exercise using the canonical commutation relation  (CCR).
\medskip
We begin with the obvious commutation relation  $[\tilde A,\tilde C] = 2b$.   Define
$$
{\Cal S}_0  =  \text{span}\{z^ke^{-(b/4)|z|^2} :  k = 0, 1, 2, \dots \}.  
$$
\noindent
Then  $\tilde A{\Cal S}_0 = \{0\}$.   For each  $j \in {\bold N}$,  define
$$
{\Cal S}_j  =  \tilde C^j{\Cal S}_0.
$$
\noindent
By the ``product rule"  for commutators, the relation  $[\tilde A,\tilde C] = 2b$  implies  
$[\tilde A,\tilde C^j] = 2bj\tilde C^{j-1}$  for  $j \geq 1$.
Let   $\varphi  \in {\Cal S}_j$  for some  $j \geq 0$.  Then there is a  $\psi \in {\Cal S}_0$  such that  
$\varphi = \tilde C^j\psi $.  Therefore  
$$
(H_b - b)\varphi =  \tilde C\tilde A\varphi =  \tilde C\tilde A\tilde C^j\psi =  \tilde C[\tilde A,\tilde C^j]\psi 
=  2bj\tilde C^j\psi = 2bj\varphi .
$$
\noindent
Hence
$$
{\Cal S}_j  \subset \text{ker}(H_b - (2j+1)b) 
\tag 2.1
$$
\noindent
for every $j \geq 0$.  Since  $H_b$  is self-adjoint, this in particular means that  ${\Cal S}_i \perp {\Cal S}_j$  for  $i \neq j$.  
\medskip
For each  $j \geq 0$,  let  ${\Cal E}_j$  be the closure of  ${\Cal S}_j$  in  $L^2({\bold C},dA)$.
\medskip
Recall that the formula
$$
(U_b\psi )(z)  =  (2/b)^{1/2}\psi ((2/b)^{1/2}z)e^{|z|^2/2},  \quad    \psi  \in L^2({\bold C},dA),
$$
\noindent
defines a unitary operator that maps  $L^2({\bold C},dA)$  onto   $L^2({\bold C},d\mu )$.  
For each  $j \geq 0$,  define
$$
{\Cal F}_j  =  U_b{\Cal E}_j.
\tag 2.2
$$
\noindent
Obviously,  we have  ${\Cal F}_0 = {\Cal F}^2$,  which is the classic Fock space.  
We will see that the spaces  ${\Cal F}_j$,  $j \geq 1$,  are in fact  
{\it higher Fock spaces}.
\medskip
Let us now denote
$$
A  =  \bar \partial  \quad \text{and}  \quad  C  =  - \partial + \bar z,
$$
\noindent
which satisfy the commutation relation  $[A,C] = 1$.
It is easy to verify that  $\langle Cu,v\rangle = \langle u,Av\rangle $  for all  $u, v \in {\bold C}[z,\bar z]$.  
That is,  over the dense subspace   ${\bold C}[z,\bar z]$  of  $L^2({\bold C},d\mu )$,  $C$  and  $A$  are each-other's adjoint.
It is also easy to verify that
$$
U_b\tilde AU_b^\ast  =  \sqrt{2b}A    \quad \text{and}  \quad  U_b\tilde CU_b^\ast  =  \sqrt{2b}C.   
$$
\noindent
Consequently,   
$$
U_bH_bU_b^\ast - b  =  2bCA.
$$
\noindent
It is easy to see that  $U_b{\Cal S}_0 = {\bold C}[z]$  and that  $U_b{\Cal S}_j = C^j{\bold C}[z]$  for every  $j \geq 1$.  
Thus by an induction on the power of  $\bar z$,  we have 
$$
U_b\text{span}\{{\Cal S}_0,{\Cal S}_1,\dots ,{\Cal S}_k,\dots \} =  {\bold C}[z,\bar z].
\tag 2.3
$$
\noindent
This shows that  $\text{span}\{{\Cal S}_0,{\Cal S}_1,\dots ,{\Cal S}_k,\dots \}$  is dense in  $L^2({\bold C},dA)$.  
For each  $j \geq 0$,  let  $E_j : L^2({\bold C},dA) \rightarrow {\Cal E}_j$  be the orthogonal projection.
Combining (2.1) with  (2.3),  we obtain
$$
H_b  =  \sum _{j=0}^\infty (2j+1)bE_j.
\tag 2.4
$$ 
\noindent
This is an explicit diagonalization of the magnetic Hamiltonian  $H_b$.
\medskip
To summarize, the spectrum of   $H_b$  comprises evenly-spaced Landau levels 
$(2j+1)b$,  $j\geq 0$. The $j$-th Landau level eigenspace  ${\Cal E}_j$  is identified with the $j$-th higher Fock space 
${\Cal F}_j$  under the unitary  $U_b$. 
\medskip
For each  $j \geq 0$,  let  $P_j : L^2({\bold C},d\mu ) \rightarrow {\Cal F}_j$  be the orthogonal projection.  
Then, of course,  $P_0 = P$.    It follows from  (2.2)  that  
$$
P_j  =  U_bE_jU_b^\ast   \quad  \text{for every}  \ \  j  \geq 0.
\tag 2.5
$$
For each   $j \geq 0$  and    $p \in {\bold C}[z]$,  we have
$$
\|C^jp\|^2  =  \langle C^jp,C^jp\rangle  =  \langle  A^jC^jp,p\rangle =  j!\langle p,p\rangle = j!\|p\|^2.
$$
\noindent
Hence,  for a given  $j \geq 0$,  if we define
$$
V_ju  =  {1\over \sqrt{j!}}C^jPu     \quad  \text{for}  \ \  u  \in  {\bold C}[z,\bar z],
\tag 2.6
$$
\noindent
then    $V_j$  extends to a partial isometry  on  $L^2({\bold C},d\mu )$  such that
$$
V_j^\ast V_j  =  P = P_0   \quad \text{and}  \quad  V_jV_j^\ast =  P_j.
\tag 2.7
$$  
\medskip
Inspired by the classic Toeplitz operators  on   ${\Cal F}^2$,  on each higher Fock space  ${\Cal F}_j$,  $j \geq 1$,  
we define the  ``Toeplitz operator"  
$$
T_{f,j}h  =  P_j(fh),  \quad  h \in {\Cal F}_j.
\tag 2.8
$$
\noindent
Given any integer (the Landau level label)  $\ell \geq 0$,  we define
$$
P^{(\ell )}  =  \sum _{j=0}^\ell P_j.
$$
\noindent
We similarly define the ``Toeplitz operator"
$$
T_f^{(\ell )}h  =  P^{(\ell )}(fh),  \quad  h \in \bigoplus _{j=0}^\ell {\Cal F}_j.
$$
\noindent
That is,  $T_f^{(\ell )} = P^{(\ell )}M_fP^{(\ell )}$.   Similar to (1.4),  we have the identity
$$
[T_f^{(\ell )},T_g^{(\ell )}]  =  P^{(\ell )}[[M_f,P^{(\ell )}],[M_g,P^{(\ell )}]]
$$
\noindent
for all  $f, g \in L^\infty ({\bold C})$  and  $\ell \geq 0$.  
\medskip
Recall the relation (2.5) that binds  $E_j$  and  $P_j$.  
Thus
$$
P^{(\ell )}  =  U_b\sum _{j=0}^\ell E_jU_b^\ast =  U_bP_{\leq E}U_b^\ast
$$
when the Fermi energy  $E$  is strictly between   $(2\ell +1)b$  and  $(2\ell +3)b$.  
For a given pair of switch functions  $f$  and  $g$,  
if it happens that both products  $[M_f,P^{(\ell )}][M_g,P^{(\ell )}]$  and   $[M_g,P^{(\ell )}][M_f,P^{(\ell )}]$  
are in the trace class, then the Kubo formula for Hall conductance reads
$$
\sigma _{\text{Hall}}(P_{\leq E})  =  -i\text{tr}[T_f^{(\ell )},T_g^{(\ell )}], 
\tag 2.9
$$
\noindent
which generalizes (1.5).   We have three goals for this paper:

(i)  Prove that for a large class of switch functions,  
the  products of commutators   

$[M_f,P^{(\ell )}][M_g,P^{(\ell )}]$  and   $[M_g,P^{(\ell )}][M_f,P^{(\ell )}]$  are indeed in the trace class.

(ii)  Compute the trace of   $\text{tr}[T_f^{(\ell )},T_g^{(\ell )}]$.  

(iii)  Identify the Hall conductance in (2.9) with a Fredholm index.  

\noindent
We will see that  (i) follows from Proposition 3.7, while (ii) and (iii) are the last two identities given in Section 8.

\bigskip
\centerline{\bf 3.  Trace class products of commutators}

\medskip
The main purpose of this section is to establish Propositions 3.7 and 3.9,  
which tell us that certain products of commutators are in the trace class. 
We begin with some of the basic estimates.
\medskip
For vectors  $x, y$  in a Hilbert space  ${\Cal H}$,  the notation  $x\otimes y$  denotes the operator 
on  ${\Cal H}$  defined by the formula
$$
x\otimes yh  =  \langle h,y\rangle x,   \quad  h \in {\Cal H}.
$$
\noindent
Note that this definition of   $x\otimes y$  is the opposite 
of the usual convention adopted by physicists, but it is the standard practice in operator theory.
\medskip
Recall that for each  $z \in {\bold C}$,  the function
$$
k_z(\zeta )  =  e^{-|z|^2/2}e^{\zeta \bar z}
$$
\noindent
is the normalized reproducing kernel for the Fock space  ${\Cal F}^2$.  
For  arbitrary  
$\varphi ,\psi \in L^2({\bold C},d\mu )$,  
$$
\langle P\varphi ,\psi \rangle =  {1\over \pi }\int _{\bold C}(P\varphi )(z)\overline{(P\psi )(z)}e^{-|z|^2}dA(z)  =
{1\over \pi }\int _{\bold C}\langle \varphi ,k_z\rangle \langle k_z,\psi \rangle dA(z).
$$
\noindent
This can be rewritten as the operator identity
$$
P  =   {1\over \pi }\int _{\bold C}k_z\otimes k_zdA(z)
\tag 3.1
$$
\noindent
on   $L^2({\bold C},d\mu )$.   Define      
$$
\Gamma = \{m+in : m, n \in {\bold Z}\}   \quad \text{and}  \quad  Q  =  \{x+iy : x, y \in [0,1)\}.
$$
\noindent
Continuing with (3.1),  we have
$$
P  =   \sum _{u\in \Gamma }{1\over \pi }\int _{Q+u}k_z\otimes k_zdA(z)
=   {1\over \pi }\int _QG_zdA(z),
\tag 3.2
$$
\noindent
where
$$
G_z  =  \sum _{u\in \Gamma }k_{u+z}\otimes k_{u+z}
$$
\noindent
for every  $z \in Q$.
\medskip
Easy calculation shows that
$$
(C^jk_z)(\zeta )  =  (\bar \zeta - \bar z)^jk_z(\zeta )  \quad  \text{for all}  \ \  j \geq 0  \ \ \text{and} \ \  z \in {\bold C}.
$$
\noindent
We now define  
$$
k_z^{(j)}(\zeta ) =  (C^jk_z)(\zeta )  =  (\bar \zeta - \bar z)^jk_z(\zeta ),
\tag 3.3
$$
\noindent
$j \geq 0$  and  $z \in {\bold C}$.  From (2.7) and (2.6)  we see that
$$
\langle  P_ju,v\rangle  =  {1\over j!}\langle C^jPA^ju,v\rangle   \quad  \text{for}  \ \  u, v \in {\bold C}[\zeta ,\bar \zeta ],
$$
\noindent
$j \geq 0$.  Combining this identity with (3.2),  we see that for each  $j \geq 0$,
$$
P_j   =   {1\over j!\pi }\int _QG_{z,j}dA(z),
\tag 3.4
$$
\noindent
where
$$
G_{z,j}  =  \sum _{u\in \Gamma }k_{u+z}^{(j)}\otimes k_{u+z}^{(j)}
\tag 3.5
$$
\noindent
for every  $z \in Q$.  Keep in mind that    $k_{u+z}^{(j)}$  is defined by (3.3). 
\medskip
For  $z \in {\bold C}$  and  $a > 0$,  denote  $B(z,a)  =  \{w \in {\bold C} : |z - w|<a\}$.  
\medskip
\noindent
{\bf Lemma 3.1.}   {\it Let}  $j \geq 0$  {\it be given}.  {\it Then there is a constant}  $0 < C_{3.1}(j) < \infty $  {\it such that for all} 
$z \in {\bold C}$   {\it and}  $\rho > 0$,  {\it we have}  $\|\chi _{{\bold C}\backslash B(z,\rho )}k_z^{(j)}\|  \leq  C_{3.1}(j)e^{-\rho ^2/3}$.
\medskip
\noindent
{\it Proof}.    Given a  $j \geq 0$,  there is a   $0 < C_1 = C_1(j) < \infty $  such that  $x^je^{-x/3} \leq C_1$  for every   $x \geq 0$.
Let  $z \in {\bold C}$   and  $\rho > 0$.   Then
$$
\align
\|\chi _{{\bold C}\backslash B(z,\rho )}k_z^{(j)}\|^2  &=  {1\over \pi }\int _{|\zeta - z|\geq \rho }|k_z^{(j)}(\zeta )|^2e^{-|\zeta |^2}dA(\zeta )  =  
{1\over \pi }\int _{|\zeta - z|\geq \rho }|\zeta - z|^{2j}e^{-|\zeta - z|^2}dA(\zeta )   \\
&\leq  {C_1\over \pi }\int _{|\zeta |\geq \rho }e^{-2|\zeta |^2/3}dA(\zeta )  
=  2C_1\int _\rho ^\infty e^{-2r^2/3}rdr  =  (3/2)C_1e^{-2\rho ^2/3}.
\endalign
$$
\noindent
Thus the constant  $C_{3.1}(j) = \{(3/2)C_1\}^{1/2}$  will do for the lemma.   $\square $
\medskip
For any nonempty subset   $F \subset {\bold C}$  and any  $z \in {\bold C}$,  denote
$$
d(z,F)  =  \inf \{|z - \zeta | : \zeta \in F\}.
$$
\noindent
For  $j \geq 0$  and  $z \in {\bold C}$,  from the facts that   $Ak_z = 0$  and  $[A,C] = 1$  we obtain
$$
\|k_z^{(j)}\|^2  =  \langle C^jk_z,C^jk_z\rangle  =  \langle A^jC^jk_z,k_z\rangle  =  j!\langle k_z,k_z\rangle = j!.
\tag 3.6
$$
\medskip
\noindent
{\bf Lemma 3.2.}    {\it Given any}  $j \geq 0$,  {\it there is a constant}  $0 < C_{3.2}(j) < \infty $  {\it such that the following estimate holds}:
{\it Let}  $\varphi \in L^\infty ({\bold C})$  {\it be such that}   $|\varphi (z)| \leq 1$  {\it for every}  $z \in {\bold C}$.
{\it Furthermore}, {\it suppose that there is a closed subset}  $F$  of  ${\bold C}$  {\it such that for every disc}  $B(w,a)$  
{\it satisfying the condition}  $B(w,a)\cap F = \emptyset $  ($w \in {\bold C}\backslash F$  {\it and}  $a > 0$),  
$\varphi $  {\it is a constant on}  $B(w,a)$.  {\it Then for every}  $z \in {\bold C}$,
$$
\|(\varphi - \varphi (z))k_z^{(j)}\|  \leq  C_{3.2}(j)\exp (-d^2(z,F)/3).
$$
\medskip
\noindent
{\it Proof}.    
Since  $\|k_z^{(j)}\| = \sqrt{j!}$  and  $\|\varphi - \varphi (z)\|_\infty \leq 2$,  
it suffices to consider  $z \in {\bold C}\backslash F$.  Let  $0 < r < d(z,F)$.  
By the assumption on  $\varphi $,  we have  $\varphi (w) = \varphi (z)$  for every  $w \in B(z,r)$.  
Therefore
$$
\|(\varphi - \varphi (z))k_z^{(j)}\|  \leq 2\|\chi _{{\bold C}\backslash B(z,r)}k_z^{(j)}\|  \leq  2C_{3.1}(j)e^{-r^2/3}
$$
\noindent
by an application of Lemma 2.1. Since this holds for every  $0 < r < d(z,F)$,  the desired conclusion follows.   $\square $
\medskip
\noindent
{\bf Lemma 3.3.}  {\it Let}     $0 < a < \infty $   {\it and define}  $V  =  \{x + iy : -a \leq x \leq a \ \text{\it and} \ y \in {\bold R}\}$.
{\it Then}  $d^2(z,i{\bold R}) \leq  2a^2 + 2d^2(z,V)$  {\it for every}  $z \in {\bold C}$.
\medskip
\noindent
{\it Proof}.  It suffices to observe that  $d(z,i{\bold R}) \leq a + d(z,V)$  for every  
$z \in {\bold C}$.    $\square $
\medskip
\noindent
{\bf Lemma 3.4.}    {\it Given any pair of}  $j \geq 0$  and  $k \geq 0$,  {\it there is a constant}  
$0 < C_{3.4}(j,k) < \infty $  {\it such that for all}  $z, w \in {\bold C}$,  
$$
\langle |k_z^{(j)}|,|k_w^{(k)}|\rangle  \leq  C_{3.4}(j,k)e^{-|z-w|^2/8}.
$$
\medskip
\noindent
{\it Proof}.  We have
$$
\align
\langle |k_z^{(j)}|,|k_w^{(k)}|\rangle  
&= {1\over \pi }\int |\zeta - z|^je^{\text{Re}(\zeta \bar z)}e^{-|z|^2/2}|\zeta - w|^ke^{\text{Re}(\zeta \bar w)}e^{-|w|^2/2}
e^{-|\zeta |^2}dA(\zeta ) \\ 
&=   {1\over \pi }\int |\zeta - z|^je^{-|\zeta - z|^2/2}|\zeta - w|^ke^{-|\zeta -w|^2/2}dA(\zeta )   \\
&= {1\over \pi }\int e^{-\{|\zeta - z|^2+|\zeta -w|^2\}/4}|\zeta - z|^j|\zeta - w|^ke^{-\{|\zeta - z|^2+|\zeta -w|^2\}/4}dA(\zeta )   \\ 
&\leq  e^{-|z-w|^2/8}{1\over \pi }\int |\zeta - z|^j|\zeta - w|^ke^{-\{|\zeta - z|^2+|\zeta -w|^2\}/4}dA(\zeta ).  
\endalign
$$
\noindent
Once   $j \geq 0$  and  $k \geq 0$  are given,  there is obviously a  $0 < C_{3.4}(j,k) < \infty $  such that 
$$
{1\over \pi }\int |\zeta - z|^j|\zeta - w|^ke^{-\{|\zeta - z|^2+|\zeta -w|^2\}/4}dA(\zeta )  \leq  C_{3.4}(j,k)
$$
\noindent
for all     $z, w \in {\bold C}$.    $\square $
\medskip
\noindent
{\bf Definition 3.5.}  Let    $0 < a < \infty $.  Then  $\Sigma _a$  denotes the collection of measurable functions  $\eta $  on  
${\bold R}$  satisfying the following three conditions:

(1)  $0 \leq \eta (x) \leq 1$  for every  $x \in {\bold R}$.

(2)  $\eta (x) = 1$  if  $x > a$.

(3)  $\eta (x) = 0$  if  $x < -a$. 
\medskip
We will use elements of   $\Sigma _a$ to construct ``switch functions'' 
on  ${\bold C}$   which interpolate between values   $0$  and $1$  within certain strips in   ${\bold C}$.
\medskip
We write  $\|\cdot \|_1$  for the norm of the trace class.
\medskip
\noindent
{\bf Lemma 3.6.}   {\it Let}  $\eta , \xi \in \Sigma _a$  {\it for some}  $0 < a < \infty $,  {\it and let}  $0 < \theta < \pi $.  
{\it Define the functions}   $f$  {\it and}  $g$  {\it on}  ${\bold C}$  {\it by the formulas}
$$
f(\zeta )  =  \eta (\text{Re}(\zeta ))  \quad \text{\it and}  \quad  g(\zeta )  =  \xi (\text{Re}(e^{-i\theta }\zeta )),  
\quad  \zeta \in {\bold C}.
$$
\noindent 
{\it Let}  $j \geq 0$  {\it and}  $k \geq 0$.  {\it Then for all}  $z, w \in Q$,  {\it the product}  $[M_f,G_{z,j}][M_g,G_{w,k}]$  
{\it is in the trace class}.  {\it Moreover},  {\it there is a}  $0 < C < \infty $  {\it which depends only on}  $a$,  $\theta $,  $j$  
{\it and}  $k$  {\it such that}
$$
\|[M_f,G_{z,j}][M_g,G_{w,k}]\|_1  \leq  C
$$ 
\noindent
{\it for every pair of}  $z, w \in Q$.
\medskip
\noindent
{\it Proof}.   Given any pair of  $z, w \in Q$,  we have
$$
[M_f,G_{z,j}][M_g,G_{w,k}]  =  \sum _{u\in \Gamma +z}
\sum _{v\in \Gamma +w}[M_f,k_u^{(j)}\otimes k_u^{(j)}][M_g,k_v^{(k)}\otimes k_v^{(k)}]  
= \sum _{u\in \Gamma +z}\sum _{v\in \Gamma +w}H_{u,v},
$$
\noindent
where
$$
\align
&H_{u,v} = \{(f-f(u))k_u^{(j)}\otimes k_u^{(j)} - k_u^{(j)}\otimes (f-f(u))k_u^{(j)}\}    \\
&\quad \quad \cdot \{(g-g(v))k_v^{(k)}\otimes k_v^{(k)} - k_v^{(k)}\otimes (g-g(v))k_v^{(k)}\}  \\
&=  \langle (g-g(v))k_v^{(k)},k_u^{(j)}\rangle (f-f(u))k_u^{(j)}\otimes k_v^{(k)}
- \langle k_v^{(k)},k_u^{(j)}\rangle (f-f(u))k_u^{(j)}\otimes (g-g(v))k_v^{(k)}  \\
&- \langle (g-g(v))k_v^{(k)},(f-f(u))k_u^{(j)}\rangle k_u^{(j)}\otimes k_v^{(k)}  
+  \langle k_v^{(k)},(f-f(u))k_u^{(j)}\rangle k_u^{(j)}\otimes (g-g(v))k_v^{(k)}.
\endalign
$$
\noindent
It suffices to estimate the sum  $\sum _{u,v}\|H_{u,v}\|_1$.
\medskip
For every pair of  $u \in \Gamma +z$  and  $v \in \Gamma + w$,  the above along with (3.6) give us
$$
\|H_{u,v}\|_1  \leq  \sqrt{k!}h_{u,v}^{(1)} +  h_{u,v}^{(2)} +  \sqrt{j!k!}h_{u,v}^{(3)} +  \sqrt{j!}h_{u,v}^{(4)},
$$
\noindent
where
$$
\align
h_{u,v}^{(1)}  &=  |\langle (g-g(v))k_v^{(k)},k_u^{(j)}\rangle |\|(f-f(u))k_u^{(j)}\|,\\
h_{u,v}^{(2)}  &=  |\langle k_v^{(k)},k_u^{(j)}\rangle |\|(f-f(u))k_u^{(j)}\|\|(g-g(v))k_v^{(k)}\|,\\
h_{u,v}^{(3)}  &=  |\langle (g-g(v))k_v^{(k)},(f-f(u))k_u^{(j)}\rangle |   \quad  \text{and}  \\
h_{u,v}^{(4)}  &=  |\langle k_v^{(k)},(f-f(u))k_u^{(j)}\rangle |\|(g-g(v))k_v^{(k)}\|.
\endalign
$$
\noindent
To prove the lemma, it suffices to find constants  $0 < C_\nu < \infty $  
such that  $\Sigma _{u,v}h_{u,v}^{(\nu )} \leq C_\nu $  for  $\nu = 1, 2, 3, 4$.
Below we present the details of the estimates for the sum  $\Sigma _{u,v}h_{u,v}^{(1)}$;  the other three sums can be handled similarly.
\medskip
First, note that since
$$
|\langle (g-g(v))k_v^{(k)},k_u^{(j)}\rangle |  
=   |\langle (g-g(v))k_v^{(k)},k_u^{(j)}\rangle |^{1/2}|\langle (g-g(v))k_v^{(k)},k_u^{(j)}\rangle |^{1/2},
$$
\noindent
we have
$$
h_{u,v}^{(1)}  \leq  \langle |k_v^{(k)}|,|k_u^{(j)}|\rangle ^{1/2}\sqrt{j!}\|(g-g(v))k_v^{(k)}\|^{1/2}\|(f-f(u))k_u^{(j)}\|.
$$
\noindent
Applying Lemmas 3.4, 3.2 and  3.3,  we obtain
$$
\align
h_{u,v}^{(1)}  &\leq C_1\big(e^{-|u-v|^2/8}\big)^{1/2}\big(e^{a^2/3}e^{-d^2(u,i{\bold R})/6}\big)^{1/2}
e^{a^2/3}e^{-d^2(v,ie^{i\theta }{\bold R})/6}   \\
&\leq C_2\exp \bigg(-{1\over 16}\left\{|u-v|^2 + d^2(u,i{\bold R}) + d^2(v,ie^{i\theta }{\bold R})\right\}\bigg).  
\tag 3.7
\endalign
$$
\noindent  
For  $x \geq 0$  and  $y \geq 0$,  $x^2 + y^2 \geq (1/2)(x+y)^2$.  By this and the triangle inequality,   
$$
\align
(1/2)|u-v|^2 + (1/2)d^2(v,ie^{i\theta }{\bold R}) &\geq (1/4)d^2(u,ie^{i\theta }{\bold R})  \quad \text{and}  \\ 
(1/2)|u-v|^2 + (1/2)d^2(u,i{\bold R}) &\geq (1/4)d^2(v,i{\bold R}).   
\endalign
$$
\noindent
Therefore from (3.7)  we deduce
$$
h_{u,v}^{(1)}  \leq C_2\exp \bigg(-{1\over 64}\left\{d^2(u,i{\bold R}) +  d^2(u,ie^{i\theta }{\bold R}) + d^2(v,i{\bold R})
+ d^2(v,ie^{i\theta }{\bold R})\right\}\bigg).
\tag 3.8
$$
\noindent
Denote  $\alpha = \min \{\sin (\theta /2),\sin((\pi -\theta )/2)\}$.  By simple plane geometry,  for every  
$q \in {\bold C}$,   
$$
\max \{d(q,i{\bold R}),d(q,ie^{i\theta }{\bold R})\}  \geq \alpha |q|.
$$
\noindent
Thus if we write  $\beta = \alpha ^2/64$,  then from (3.8)   we obtain
$$
h_{u,v}^{(1)}  \leq C_2e^{-\beta |u|^2}e^{-\beta |v|^2}.
$$
\noindent
For  $x \in \Gamma $,  $|z|^2 + |x+z|^2 \geq (1/2)|x|^2$.  Therefore
$$
\align
\sum _{u\in \Gamma +z}&\sum _{v\in \Gamma +w}h_{u,v}^{(1)}  
\leq  C_2\sum _{u\in \Gamma +z}e^{-\beta |u|^2}\sum _{v\in \Gamma +w}e^{-\beta |v|^2}  \\
&\leq  C_2\sum _{x\in \Gamma }e^{-(\beta /2)|x|^2+|z|^2}\sum _{y\in \Gamma }e^{-(\beta /2)|y|^2+|w|^2}  
\leq e^4C_2\bigg(\sum _{x\in \Gamma }e^{-(\beta /2)|x|^2}\bigg)^2.
\endalign
$$
\noindent
This completes the proof.   $\square $ 
\medskip
\noindent
{\bf Proposition 3.7.}  {\it Let}  $\eta , \xi \in \Sigma _a$  {\it for some}  $0 < a < \infty $.  {\it Let}  $0 < \theta < \pi $.  
{\it Define the functions}   $f$  {\it and}  $g$  {\it on}  ${\bold C}$  {\it by the formulas}
$$
f(\zeta )  =  \eta (\text{Re}(\zeta ))  \quad \text{\it and}  \quad  g(\zeta )  =  \xi (\text{Re}(e^{-i\theta }\zeta )),
$$
\noindent 
$\zeta  \in {\bold C}$.  {\it Then for all}  $j \geq 0$  {\it and}  $k \geq 0$,  {\it the operator}  $[M_f,P_j][M_g,P_k]$  {\it is in the trace class}.  
{\it Consequently},  {\it for every}  $\ell \geq 0$,  {\it the commutator}  $[T_f^{(\ell )},T_g^{(\ell )}]$  {\it is in the trace class}. 
\medskip
\noindent
{\it Proof}.   It follows from (3.4) that 
$$
[M_f,P_j][M_g,P_k]  =  {1\over j!k!\pi ^2}\iint _{Q\times Q}[M_f,G_{z,j}][M_g,G_{w,k}]dA(z)dA(w).
$$
\noindent
Therefore
$$
\|[M_f,P_j][M_g,P_k]\|_1  \leq {1\over j!k!\pi ^2}\iint _{Q\times Q}\|[M_f,G_{z,j}][M_g,G_{w,k}]\|_1dA(z)dA(w).
$$
\noindent
By Lemma 3.6, the right-hand side is finite.  $\square $
\medskip
In terms of simple plane geometry, we can think of the commutators   
$[M_f,P_j]$  and  $[M_g,P_k]$  in  Proposition 3.7  as being ``supported"  in strips 
at the angle  $\theta $  to each other.  The intersection of the two strips is a parallelogram,  which is why the product  
$[M_f,P_j][M_g,P_k]$  is in the trace class.  Proposition 3.9 below uses a variant of this idea.
\medskip
\noindent
{\bf Lemma 3.8.}    {\it Let}  $s, t \in {\bold R}$  {\it be such that}  $s < t < s + \pi $.  {\it Define the subset}  
$$
W = \{re^{ix} :  s \leq x \leq t  \ \text{and} \ r \geq 0\}
$$
\noindent
{\it of}  ${\bold C}$.  {\it Let}   $\theta  \in {\bold R}$  {\it satisfy the conditions}  $ie^{i\theta }{\bold R}\cap e^{is}{\bold R} = \{0\}$  {\it and}  
$ie^{i\theta }{\bold R}\cap e^{it}{\bold R} = \{0\}$
{\it (see Figure 1)}.  
 {\it Pick a}  $\xi \in \Sigma _a$  {\it for some}  $0 < a < \infty $  {\it and define the function}
$$
g(\zeta )  =  \xi (\text{Re}(e^{-i\theta }\zeta )),
$$
\noindent
$\zeta  \in {\bold C}$.  {\it Given any}  $j \geq 0$  {\it and}  $k \geq 0$,   {\it there is a}  $0 < C < \infty $  {\it such that}
$$
\|[M_{\chi _W},G_{z,j}][M_g,G_{w,k}]\|_1  \leq  C
$$
\noindent
{\it for every pair of}  $z, w \in Q$.
\medskip
\noindent
{\it Proof}.   Similar to what happened in the proof of Lemma 3.6,  for any $z, w \in Q$,  we have
$$
\align
[M_{\chi _W},G_{z,j}][M_g,G_{w,k}]  
&=  \sum _{u\in \Gamma +z}\sum _{v\in \Gamma +w}[M_{\chi _W},k_u^{(j)}\otimes k_u^{(j)}][M_g,k_v^{(k)}\otimes k_v^{(k)}]  \\
&= \sum _{u\in \Gamma +z}\sum _{v\in \Gamma +w}H_{u,v},
\endalign
$$
\noindent
where
$$
\align
H_{u,v}  
&=  \langle (g-g(v))k_v^{(k)},k_u^{(j)}\rangle (\chi _W-\chi _W(u))k_u^{(j)}\otimes k_v^{(k)}   \\
&- \langle k_v^{(k)},k_u^{(j)}\rangle (\chi _W-\chi _W(u))k_u^{(j)}\otimes (g-g(v))k_v^{(k)}   \\
&- \langle (g-g(v))k_v^{(k)},(\chi _W-\chi _W(u))k_u^{(j)}\rangle k_u^{(j)}\otimes k_v^{(k)}    \\
&+ \langle k_v^{(k)},(\chi _W-\chi _W(u))k_u^{(j)}\rangle k_u^{(j)}\otimes (g-g(v))k_v^{(k)}.
\endalign
$$
\noindent
Again, it suffices to estimate the sum  $\sum _{u,v}\|H_{u,v}\|_1$.
\medskip
For every pair of  $u \in \Gamma +z$  and  $v \in \Gamma + w$,   
$$
\|H_{u,v}\|_1  \leq  \sqrt{k!}h_{u,v}^{(1)} +  h_{u,v}^{(2)} +  \sqrt{j!k!}h_{u,v}^{(3)} +  \sqrt{j!}h_{u,v}^{(4)},
$$
\noindent
where
$$
\align
h_{u,v}^{(1)}  &=  |\langle (g-g(v))k_v^{(k)},k_u^{(j)}\rangle |\|(\chi _W-\chi _W(u))k_u^{(j)}\|,\\
h_{u,v}^{(2)}  &=  |\langle k_v^{(k)},k_u^{(j)}\rangle |\|(\chi _W-\chi _W(u))k_u^{(j)}\|\|(g-g(v))k_v^{(k)}\|,\\
h_{u,v}^{(3)}  &=  |\langle (g-g(v))k_v^{(k)},(\chi _W-\chi _W(u))k_u^{(j)}\rangle |   \quad  \text{and}  \\
h_{u,v}^{(4)}  &=  |\langle k_v^{(k)},(\chi _W-\chi _W(u))k_u^{(j)}\rangle |\|(g-g(v))k_v^{(k)}\|.
\endalign
$$
\noindent
This time, let us estimate    $\Sigma _{u,v}h_{u,v}^{(3)}$;  the other three sums are handled similarly.
\medskip
It is easy to see that
$$
h_{u,v}^{(3)} \leq  \langle |k_v^{(k)}|,|k_u^{(j)}|\rangle ^{1/2}\|(g-g(v))k_v^{(k)}\|^{1/2}\|(\chi _W-\chi _W(u))k_u^{(j)}\|^{1/2}.
$$
\noindent
Applying Lemmas 3.4, 3.2 and 3.3,  we obtain
$$
h_{u,v}^{(3)}  \leq C_1\big(e^{-|u-v|^2/8}\cdot e^{a^2/3}e^{-d^2(v,ie^{i\theta }{\bold R})/6}\cdot 
e^{-d^2(u,\partial W)/3}\big)^{/1/2}. 
$$
\noindent
Obviously,  $\partial W \subset e^{is}{\bold R}\cup e^{it}{\bold R}$.  Therefore the above implies
$$
h_{u,v}^{(3)}  \leq C_2\max \{a_{u,v},b_{u,v}\} \leq  C_2(a_{u,v} + b_{u,v}),  
$$
\noindent
where 
$$
\align 
a_{u,v}  &=       \exp \bigg(-{1\over 16}\left\{|u-v|^2 + d^2(v,ie^{i\theta }{\bold R}) + d^2(u,e^{is}{\bold R})\right\}\bigg)  \quad  \text{and}   \\
b_{u,v}  &=       \exp \bigg(-{1\over 16}\left\{|u-v|^2 + d^2(v,ie^{i\theta }{\bold R}) + d^2(u,e^{it}{\bold R})\right\}\bigg).
\endalign
$$
\noindent
Let us estimate    $\Sigma _{u,v}b_{u,v}$;  the sum  $\Sigma _{u,v}a_{u,v}$  is handled similarly.
\medskip
By the reasoning before inequality (3.8),  we have
$$
b_{u,v}  \leq  \exp \bigg(-{1\over 64}\left\{d^2(v,e^{it}{\bold R}) +  d^2(v,ie^{i\theta }{\bold R}) + d^2(u,e^{it}{\bold R})
+ d^2(u,ie^{i\theta }{\bold R})\right\}\bigg).
\tag 3.9
$$
\noindent
Let  $\gamma $  be the angle between the lines  $e^{it}{\bold R}$  and  $ie^{i\theta }{\bold R}$.   By definition,  
$0 \leq \gamma \leq \pi $.  By our assumption on  $t$ and $\theta $,  $\gamma \neq 0$  and  $\gamma \neq \pi $.
For  each  $q \in {\bold C}$,  plane geometry gives us the inequality 
$$
\max \{d(q,e^{it}{\bold R}),d(q,ie^{i\theta }{\bold R})\}  \geq  \alpha |q|,
$$
\noindent
where  $\alpha = \min \{\sin (\gamma /2),\sin ((\pi - \gamma )/2)\}$. 
Thus if we write  $\beta = \alpha ^2/64$,  then from (3.9)   we obtain
$$
b_{u,v} \leq e^{-\beta |u|^2}e^{-\beta |v|^2}.
$$
\noindent
By the argument at the end of the proof of Lemma 2.6,  we have
$$
\sum _{u\in \Gamma +z}\sum _{v\in \Gamma +w}b_{u,v}   
\leq e^4\bigg(\sum _{x\in \Gamma }e^{-(\beta /2)|x|^2}\bigg)^2.
$$
\noindent
This completes the proof.   $\square $
\medskip
\noindent
{\bf Proposition 3.9.}    {\it Let}  $s, t \in {\bold R}$  {\it be such that}  $s < t < s + \pi $.  {\it Define the subset}  
$$
W = \{re^{ix} :  s \leq x \leq t  \ \text{\it and} \ r \geq 0\}
$$
\noindent
{\it of}  ${\bold C}$.  {\it Let}   $\theta  \in {\bold R}$  {\it satisfy the conditions}  $ie^{i\theta }{\bold R}\cap e^{is}{\bold R} = \{0\}$ {\it and}  
$ie^{i\theta }{\bold R}\cap e^{it}{\bold R} = \{0\}$.  {\it Pick a}  $\xi \in \Sigma _a$  {\it for some}  $0 < a < \infty $  {\it and define the function}
$$
g(\zeta )  =  \xi (\text{Re}(e^{-i\theta }\zeta )),
$$
\noindent
$\zeta  \in {\bold C}$.  {\it Then for all}  $j \geq 0$  {\it and}  $k \geq 0$,  {\it the operator}   $[M_{\chi _W},P_j][M_g,P_k]$  
{\it is in the trace class}.   {\it Consequently},  {\it for every}  $\ell \geq 0$,  {\it the semi-commutators}  
$$
T_{\chi _W}^{(\ell )}T_g^{(\ell )} - T_{\chi _Wg}^{(\ell )}   \quad  \text{\it and}  \quad   
T_g^{(\ell )}T_{\chi _W}^{(\ell )} - T_{g\chi _W}^{(\ell )}
$$  
\noindent
{\it are in the trace class}.   
\medskip
\noindent
{\it Proof}.   This proposition is derived from Lemma 3.8 in exactly the same way Proposition 3.7 
was derived from Lemma 3.6.  $\square $
\medskip
\noindent
{\bf Proposition 3.10.}  {\it Let}  $\varphi \in L^\infty ({\bold C})$.  {\it If the support of}  $\varphi $  {\it is contained in a} 
{\it bounded set},  {\it then for every}  $j \geq 0$,  {\it the operator}  $M_\varphi P_j$  {\it is in the trace class}.
\medskip
\noindent
{\it Proof}.  Let  a  $j \geq 0$  be given.  By (3.4),  it suffices to find a  $0 < C < \infty $  such that
$$
\|M_\varphi G_{z,j}\|_1  \leq  C   \quad \text{for every}  \ \  z \in Q.
\tag 3.10  
$$ 
\noindent
By (3.5),  for each  $z \in Q$  we have
$$
\|M_\varphi G_{z,j}\|_1 \leq  \sqrt{j!}\sum _{u\in \Gamma }\|\varphi k_{u+z}^{(j)}\|
\tag 3.11
$$
\noindent
By assumption,  there is a  $0 < \rho < \infty $  such that  $\varphi = 0$  on  
${\bold C}\backslash B(0,\rho )$.  Thus if  $|u| > 2\rho + 4$,  then  $B(u+z,|u|/2)\cap B(0,\rho ) = \emptyset $,  
i.e.,  $B(0,\rho ) \subset {\bold C}\backslash B(u+z,|u|/2)$,  and we have
$$
\|\varphi k_{u+z}^{(j)}\|  \leq  \|\varphi \|_\infty \|\chi _{{\bold C}\backslash B(u+z,|u|/2)}k_{u+z}^{(j)}\|
\leq  C_{3.2}(j) \|\varphi \|_\infty e^{-|u|^2/12},
$$
\noindent
where the second  $\leq $  follows from Lemma 3.1.  Thus
$$
\sum _{u\in \Gamma \backslash B(0,2\rho +5)}\|\varphi k_{u+z}^{(j)}\| 
\leq  C_{3.2}(j)\|\varphi \|_\infty \sum _{u\in \Gamma \backslash B(0,2\rho +5)}e^{-|u|^2/6}.
$$
\noindent
Combining this bound with (3.11),  we obtain (3.10).  
$\square $ 

\midinsert
\epsfxsize=0.6\hsize                      
\centerline{\epsfbox{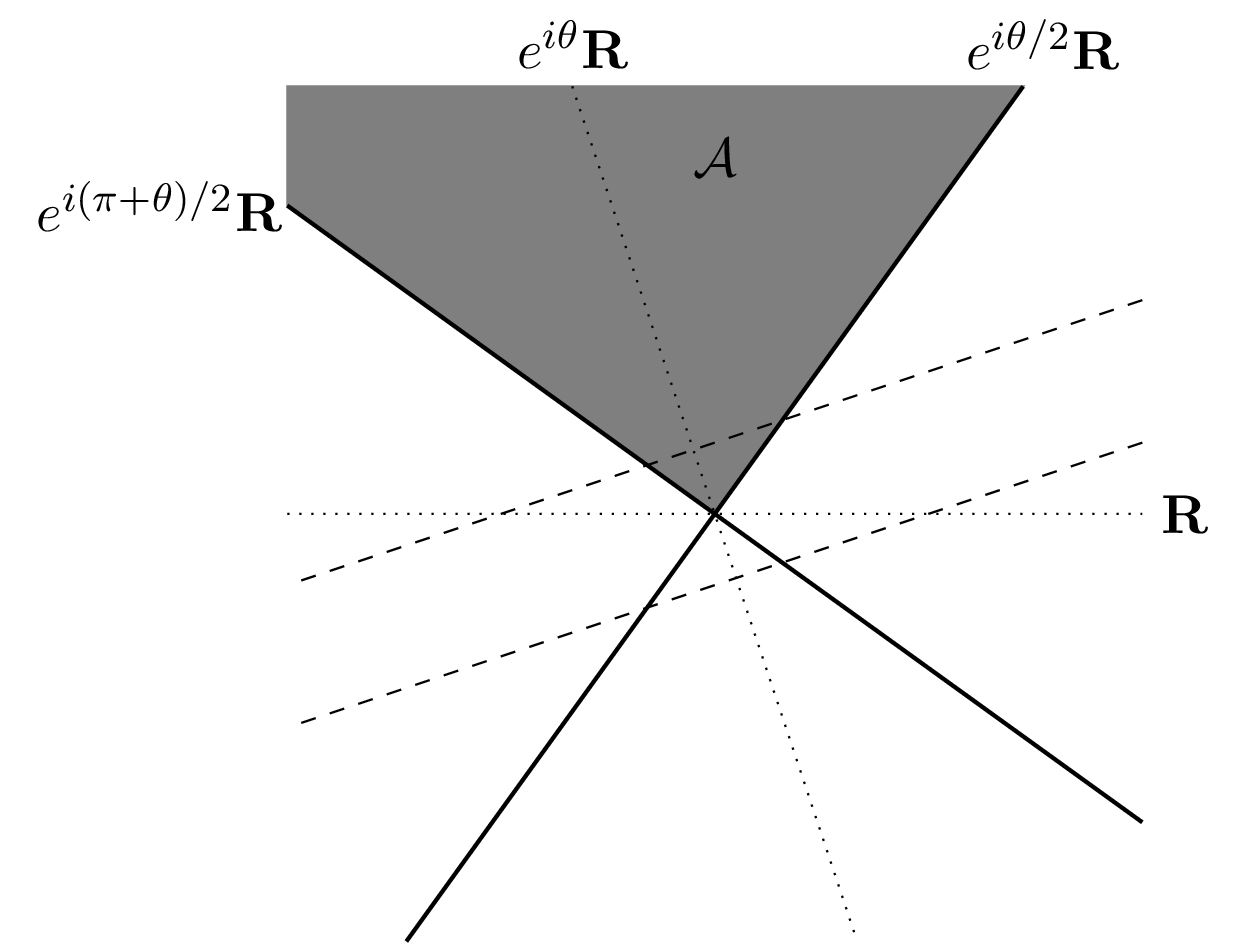}}   
\captionwidth{0.9\hsize}
\botcaption{Figure 1} For $0<\theta<\pi$, the wedge ${\Cal A}$ (shaded region) with extremal rays at angles $s=\theta/2$ and $t=(\pi+\theta)/2$ satisfies the condition of Lemma 3.8 and Proposition 3.9. Furthermore, strips perpendicular to $e^{i\theta}{\bold R}$ intersect ${\Cal A}$ on a compact set (triangle). 
\endcaption
\vspace{1em}
\endinsert

\bigskip
\centerline{\bf 4. Fredholm inverse}
\medskip
In this section we deal with Toeplitz operators whose Fredholm inverse cannot be constructed by known methods 
in the literature.  These Toeplitz operators have symbols that oscillate at infinity.
\medskip
Take an  $a > 0$  and pick  $\eta , \xi \in \Sigma _a$.  Also, pick a  $0 < \theta < \pi $.  
We now fix the switch functions
$$
f_1(\zeta )  =  \eta (\text{Re}(\zeta ))  \quad \text{and}  \quad  f_2(\zeta )  =  \xi (\text{Re}(e^{-i\theta }\zeta )),  
\tag 4.1
$$
\noindent 
$\zeta  \in {\bold C}$. Furthermore,  we define the function   
$$
F = f_1 + if_2
$$ 
\noindent
on  ${\bold C}$.   We remind the reader that  
$$
S  =  \{x+iy : x, y \in [0,1]\}.
$$
\noindent
Recall that   (2.8)  defines  the  ``Toeplitz operator"    $T_{f,j}$  on  ${\Cal F}_j$,  $j \geq 0$.
\medskip
\noindent
{\bf Theorem  4.1.}  {\it On the space}  ${\Cal F}_j$,  $j \geq 0$,  {\it the essential spectrum of the Toeplitz operator}  $T_{F,j}$   
{\it is contained in}  $\partial S$,  {\it the boundary of the square}  $S$.  
\medskip
To prove Theorem 4.1, we first establish a general ``product rule" for the spectra of self-adjoint operators:  
\medskip
\noindent
{\bf Lemma 4.2.}   {\it Let}  $A$,  $B$  {\it be bounded self-adjoint operators on a Hilbert space}  ${\Cal H}$.  
{\it Suppose that the spectra of}  $A$  {\it and}  $B$  {\it are contained in the intervals}  $[a,b]$  {\it and}  $[c,d]$  
{\it respectively}.   {\it Then the spectrum of}  $T = A + iB$  {\it is contained in the rectangle}
$$
\{x + iy :  x \in [a,b], y \in [c,d]\}.
$$
\medskip
\noindent
{\it Proof}.   Let  $t \in {\bold R}\backslash [a,b]$  and  $y \in {\bold R}$.  Let us show that  $T - (t+iy)$  is invertible.  
If  $t < a$,  then  $A - t > 0$,  and  $(A - t)^{1/2}$  has a bounded inverse   $(A - t)^{-1/2}$.  In this case,
$$
T - (t+iy)  =  (A - t)^{1/2}\{1 + i(A - t)^{-1/2}(B-y)(A - t)^{-1/2}\}(A - t)^{1/2}.
$$
\noindent
Since the operator  $(A - t)^{-1/2}(B-y)(A - t)^{-1/2}$  is self-adjoint,  the above is invertible.   In the case  $t > b$,  
we have  $t - A > 0$  and  
$$
T - (t+iy)  =  (t - A)^{1/2}\{-1 + i(t - A)^{-1/2}(B-y)(t - A)^{-1/2}\}(t - A)^{1/2},
$$
\noindent
which is also invertible.   A similar argument shows that if  $x \in {\bold R}$  and  $t \in {\bold R}\backslash [c,d]$,  then  
$T - (x+it)$  is invertible.    $\square $
\noindent
\medskip
\noindent
{\it Proof of Theorem} 4.1. By Lemma 4.2, it suffices to show that the interior of  $S$  
does not intersect the essential spectrum of  $T_{F,j}$.   We begin with the following four wedges in  ${\bold C}$:
$$
\align
{\Cal A}  &=   \{re^{ix} : \theta /2 \leq x \leq (\pi +\theta )/2 \ \text{and} \ r \geq 0\},  \\
{\Cal B}  &=   \{re^{ix} : (\pi +\theta )/2 \leq x \leq \pi +(\theta /2) \ \text{and} \ r \geq 0\},  \\
{\Cal C}  &=   \{re^{ix} : \pi +(\theta /2) \leq x \leq (3\pi +\theta )/2 \ \text{and} \ r \geq 0\}    \quad \text{and}  \\
{\Cal D}  &=   \{re^{ix} :(3\pi +\theta )/2 \leq x \leq 2\pi + (\theta /2)  \ \text{and} \ r \geq 0\},
\endalign
$$
\noindent
see Figure 1.  
\noindent
Thus  ${\Cal B} = i{\Cal A}$,  ${\Cal C} = -{\Cal A}$,  and  ${\Cal D} = -i{\Cal A}$.  We claim that
$$
\align
T_{f_2,j}T_{\chi _{\Cal A},j}  &=   T_{\chi _{\Cal A},j}  + K_{\Cal A},  
\tag 4.2  \\
T_{f_1,j}T_{\chi _{\Cal B},j}  &=     K_{\Cal B}, 
\tag 4.3   \\
T_{f_2,j}T_{\chi _{\Cal C},j}  &=   K_{\Cal C}  \quad  \text{and} 
\tag 4.4 \\
T_{f_1,j}T_{\chi _{\Cal D},j}  &=   T_{\chi _{\Cal D},j}  + K_{\Cal D},  
\tag 4.5  \\
\endalign
$$
\noindent
where   $K_{\Cal A}$,  $K_{\Cal B}$,  $K_{\Cal C}$,  $K_{\Cal D}$  are compact operators.   Let us consider (4.2).
\medskip
By Proposition 3.9, we have  $T_{f_2,j}T_{\chi _{\Cal A},j}  =   T_{f_2\chi _{\Cal A},j}  + K_{\Cal A}^{(1)}$,  
where  $K_{\Cal A}^{(1)}$  is compact.  
Thus
$$
T_{f_2,j}T_{\chi _{\Cal A},j}  =   T_{\chi _{\Cal A},j} - T_{(1-f_2)\chi _{\Cal A},j}  + K_{\Cal A}^{(1)}.
\tag 4.6
$$
\noindent
By the definition of  $f_2$, if  $\zeta = e^{i\theta }(x+iy)$  for  $x > a$  and  $y \in {\bold R}$,  then  $f_2(\zeta ) = 1$.  
Thus  the support of the function   $(1-f_2)\chi _{\Cal A}$  is contained in the set
$$
{\Cal A}\cap \{e^{i\theta }(x+iy) : x \leq a \  \text{and}  \  y \in {\bold R}\}.
$$
\noindent
Obviously, this is the region enclosed by the lines  $e^{i\theta /2}{\bold R}$,  $e^{i(\pi +\theta )/2}{\bold R}$  and  
$\{e^{i\theta }(a+iy) : y \in {\bold R}\}$,  which is a triangle
(see Figure 1).  
 By Proposition 3.10, the Toeplitz operator  
$T_{(1-f_2)\chi _{\Cal A},j}$  is compact.  Therefore (4.2) follows from (4.6).  The proofs of (4.3), (4.4) and (4.5) are similar 
and will be omitted.  
\medskip
Let  $\lambda \in S\backslash \partial S$.  That is,
$$
\lambda  =  \alpha + i\beta ,  \quad \text{where} \ \ \alpha , \beta \in (0,1). 
$$
\noindent
Since the operators  $T_{f_1,j}$  and  $T_{f_2,j}$  are self-adjoint,  the operators  
$$
T_{f_1,j} - \alpha + i(1-\beta ),  \quad T_{if_2,j} - \alpha - i\beta ,  \quad  T_{f_1,j} - \alpha - i\beta \quad \text{and} \quad  
T_{if_2,j} + (1-\alpha ) - i\beta 
$$
\noindent
are invertible on  ${\Cal F}_j$.     Let  $A$, $B$,  $C$,  $D$  denote their respective inverses.  Then
$$
\align
(T_{f_1,j} - \alpha + i(1-\beta ))A &= 1,  \\  
(T_{if_2,j} - \alpha - i\beta )B &=1,  \\  
(T_{f_1,j} - \alpha - i\beta )C &= 1  \quad  \text{and}  \\
(T_{if_2,j} + (1-\alpha ) - i\beta )D &= 1.
\endalign
$$
\noindent
Combining these identities with (4.2)-(4.5),  and with the fact that the commutators  $[T_{f_1,j},T_{\chi _{\Cal E},j}]$  and  
$[T_{f_2,j},T_{\chi _{\Cal E},j}]$  are in the trace class for every  ${\Cal E} \in \{{\Cal A}, {\Cal B}, {\Cal C}, {\Cal D}\}$   
(see (1.4) and Proposition 3.9),   we find that
$$
\align
(T_{F,j} - \lambda )(T_{\chi _{\Cal A},j}A + T_{\chi _{\Cal B},j}B + T_{\chi _{\Cal C},j}C + T_{\chi _{\Cal D},j}D)  
&=  T_{\chi _{\Cal A},j} + T_{\chi _{\Cal B},j} + T_{\chi _{\Cal C},j} + T_{\chi _{\Cal D},j} + K  \\
&=  1 + K,
\endalign
$$
\noindent
where  $K$  is a compact operator.  Similarly,  we have
$$
\align
(AT_{\chi _{\Cal A},j} + BT_{\chi _{\Cal B},j} + CT_{\chi _{\Cal C},j} + DT_{\chi _{\Cal D},j})(T_{F,j} - \lambda ) 
&= T_{\chi _{\Cal A},j} + T_{\chi _{\Cal B},j} + T_{\chi _{\Cal C},j} + T_{\chi _{\Cal D},j} + L   \\   
&= 1 + L,
\endalign
$$
\noindent
where  $L$  is a compact operator.  The above two identities show that  $\lambda $  is not in 
the essential spectrum of  $T_F$.   This completes the proof.    $\square $
\medskip
\noindent
{\bf Theorem  4.3.}  {\it Let}  $\ell \geq 0$.  {\it Then on the  space} 
${\Cal F}_0\oplus \cdots \oplus {\Cal F}_\ell $,  {\it the essential spectrum of the Toeplitz operator}  
$T_F^{(\ell )}$   {\it is contained in}  $\partial S$,  {\it the boundary of the square}  $S$.  
\medskip
\noindent
{\it Proof}.   It follows from Proposition 3.7 that the commutator  $[T_{f_1}^{(\ell )},T_{f_2}^{(\ell )}]$  is in the trace class.  
Therefore by the argument in part (1) of the proof of Theorem 4.1, the essential spectrum of  $T_F^{(\ell )}$   is contained in   $S$.  
\medskip
It also follows from   Proposition 3.7  that the operators   $T_{f_i}^{(\ell )}T_{\chi _{\Cal E}}^{(\ell )} - T_{f_i\chi _{\Cal E}}^{(\ell )}$  
and  $T_{\chi _{\Cal E}}^{(\ell )}T_{f_i}^{(\ell )} - T_{\chi _{\Cal E}f_i}^{(\ell )}$  are in the trace class for all   $i \in \{1,2\}$  
and  ${\Cal E} \in \{{\Cal A},{\Cal B},{\Cal C},{\Cal D}\}$,  where the wedges    ${\Cal A}$, ${\Cal B}$, ${\Cal C}$,  ${\Cal D}$  
were defined in the previous proof.  Therefore the argument in part (2) of the proof of Theorem 4.1 also works here.  By that argument,  
if  $\lambda \in S\backslash \partial S$,  then  $\lambda $  is not in the essential spectrum of   $T_F^{(\ell )}$.  
This completes the proof.   $\square $

\bigskip
\centerline{\bf 5. Trace calculation}
\medskip
In this section we revisit the trace calculation, equation (1.6), carried out  in [11].  
Thus in this section we only consider the classic Fock space  ${\Cal F}^2 = {\Cal F}_0$  and classic 
Toeplitz operators on it.  The calculation presented below 
differs slightly from the calculation in [11] in the respect that we allow any  $0 < \theta < \pi $  in (4.1).  
Because of this, the trace calculation requires one extra cancellation argument.
\medskip
As observed in [1, Proposition 6.9],  for all  $\Lambda \in \Sigma _a$  and  $t \in {\bold R}$,  we have
$$
\int _{\bold R}(\Lambda (x+t) - \Lambda (x))dx = t.
\tag 5.1
$$
\medskip
\noindent
{\bf Proposition 5.1.}  {\it For the switch functions}  $f_1$, $f_2$  {\it defined by}  (4.1), {\it we have}
$$
\text{tr}[T_{f_1},T_{f_2}]  =  {1\over 2\pi i}.
$$
\medskip
\noindent
{\it Proof}.  By (1.4),  we have
$$
[T_{f_1},T_{f_2}]  = P[M_{f_1},P][M_{f_2},P]  -  P[M_{f_2},P][M_{f_1},P].  
\tag 5.2
$$
\noindent
By Proposition 3.7, both terms on the right-hand side are in the trace class, 
which is a fact that we will use in the calculation.
\medskip
We can regard  $P$  as an integral operator on  $L^2({\bold C},d\mu )$  with the function  $e^{\zeta \bar z}$  
as its integral kernel.  Thus  $P[M_{f_1},P][M_{f_2},P]$  and  $P[M_{f_2},P][M_{f_1},P]$
are integral operators on  $L^2({\bold C},d\mu )$  with the functions
$$
\align
K_{1,2}(\zeta ,z)  &=  \iint e^{\zeta  \bar w}e^{w \bar u}e^{u \bar z}
(f_1(w) - f_1(u))(f_2(u) - f_2(z))d\mu (w)d\mu (u)  \quad \text{and}  \\
K_{2,1}(\zeta ,z)  &=  \iint e^{\zeta  \bar w}e^{w \bar u}e^{u \bar z}
(f_2(w) - f_2(u))(f_1(u) - f_1(z))d\mu (w)d\mu (u) 
\endalign
$$
\noindent
as their respective integral kernels.   Therefore
$$
\align
\text{tr}(P[M_{f_1},P]&[M_{f_2},P])  =  \int K_{1,2}(\zeta ,\zeta )d\mu (\zeta )  \\
&= \iiint e^{\zeta  \bar w}e^{w \bar u}e^{u \bar \zeta }
(f_1(w) - f_1(u))(f_2(u) - f_2(\zeta ))d\mu (w)d\mu (u)d\mu (\zeta ).
\endalign
$$
\noindent
It is easy to verify that the identity
$$
(\zeta \bar w + w\bar u + u\bar \zeta ) - (|\zeta |^2 + |w|^2 + |u|^2)  =  (\zeta - u)\overline{(w-u)}  -  |\zeta - u|^2 - |w - u|^2
$$
\noindent
holds for  all  $\zeta , w, u \in {\bold C}$.  Therefore
$$
\align
\text{tr}&(P[M_{f_1},P][M_{f_2},P]) \\ 
&=  {1\over \pi ^3}\iiint e^{(\zeta - u)\overline{(w-u)}}e^{- |\zeta - u|^2 - |w - u|^2}
(f_1(w) - f_1(u))(f_2(u) - f_2(\zeta ))dA(w)dA(u)dA(\zeta )   \\
&=  {1\over \pi ^3}\iint e^{x\bar y}e^{- |x|^2 - |y|^2}
\bigg\{\int (f_1(u+y) - f_1(u))(f_2(u) - f_2(u+x))dA(u)\bigg\}dA(x)dA(y).
\endalign
$$
\noindent
Let us write  $x = x_1 +ix_2$  and  $y = y_1 + iy_2$.   
Combining (5.1) with easy plane geometry,  we find that  
$$
\int (f_1(u+y) - f_1(u))(f_2(u) - f_2(u+x))dA(u)  =  -y_1(x_2 + \cot \theta x_1).
\tag 5.3
$$
\noindent
(Those who need the details, see the Remark below.)  Hence
$$
\text{tr}(P[M_{f_1},P][M_{f_2},P])  =   {-1\over \pi ^3}\iint e^{x\bar y}e^{- |x|^2 - |y|^2}y_1(x_2 + \cot \theta x_1)dA(x)dA(y).
$$
\noindent
Similarly,
$$
\text{tr}(P[M_{f_2},P][M_{f_1},P])  =   {-1\over \pi ^3}\iint e^{x\bar y}e^{- |x|^2 - |y|^2}(y_2+\cot \theta y_1)x_1dA(x)dA(y).
$$
\noindent
Therefore, by (5.2),  
$$
\text{tr}[T_{f_1},T_{f_2}]  =  {1\over \pi ^3}\iint e^{x\bar y}e^{- |x|^2 - |y|^2}(y_2x_1 - y_1x_2)dA(x)dA(y),
$$
\noindent
which is independent of  $\theta $.  
The last step in the trace calculation is exactly the same as in [11].  That is,  
integration in polar coordinates for both variables  $x$ and $y$  yields
$$
\text{tr}[T_{f_1},T_{f_2}]  =  {1\over 2\pi i}.
$$
\noindent
$\square $
\medskip
\noindent
{\bf Remark.}  Let  $L$  be a line that intersects the  $x$-axis at the angle  $\theta $.  Let  $S_1$  be a strip perpendicular 
to the $x$-axis,  and let  $S_2$  be a strip perpendicular to  $L$.  Identity (5.3) is obtained by computing the area of the 
parallelogram   $S_1\cap S_2$.
\medskip
As we have already mentioned, Proposition 5.1 only covers the setting of the classic Fock space  ${\Cal F}^2 = {\Cal F}_0$.  
Its generalization to the higher Fock spaces  ${\Cal F}_j$,  $j \geq 1$,  requires   
the Carey-Pincus theory of principal functions, which we review next. 

\bigskip
\centerline{\bf 6.  Trace and index}
\medskip
To express the quantized Hall conductance 
in terms of a Fredholm index, in addition to the work we do in this paper,  we need the theories of  Helton-Howe [8]  and  
Carey-Pincus [3,4,10] for almost commuting pairs of self-adjoint operators.
\medskip
Suppose that  $A$,  $B$  are bounded self-adjoint operators such that the commutator  
$[A,B]$  is in the trace class.   In [8],  Helton and Howe showed that there is a compactly-supported, 
real-valued regular Borel measure  $dP$  on  ${\bold R}^2$  such that 
$$
\text{tr}([p(A,B),q(A,B)])  =  i\int \{p,q\}dP
$$ 
\noindent
for all  $p, q \in {\bold C}[x,y]$.   Here,  $\{p,q\}$  is the Poisson bracket for   $p, q$.   That is,  
$$
\{p,q\}(x,y)  =  {\partial p\over \partial x}(x,y){\partial q\over \partial y}(x,y)  
-  {\partial p\over \partial y}(x,y){\partial q\over \partial x}(x,y).
$$
\noindent
Carey and Pincus  [3,4]  took this trace formula one step further by showing that there is a boundedly-supported  
$g_{A,B} \in L^1({\bold R}^2)$,  which is called the {\it principal function}  for the pair  $A$,  $B$,  such that 
$$
\text{tr}([p(A,B),q(A,B)])  =  {-1\over 2\pi i}\iint \{p,q\}(x,y)g_{A,B}(x,y)dxdy
\tag 6.1
$$ 
\noindent
for all  $p, q \in {\bold C}[x,y]$.    In other words,  (6.1)  tells us   
that the measure   $dP$  is absolutely continuous with respect to
the two-dimensional Lebesgue measure on  ${\bold R}^2$.  
By functional calculus,  
(6.1)  extends to a much larger class of functions than   ${\bold C}[x,y]$.   
For an irreducible pair   $A$,  $B$  with $\text{rank}([A,B]) = 1$,   the principal function  $g$  is a complete unitary invariant.
\medskip
Let  $T = A + iB$.  For our purpose, the
more important fact is that  for each point  $(x,y)$  such that   $x + iy$  is not in the essential spectrum of  $T$,  
$$
g_{A,B}(x,y)  =  \text{index}(T - (x+iy)).  
\tag 6.2
$$
\noindent
See [3, Theorem 4], or [4, Theorem 8.1].
\medskip
By (1.4) and Proposition 3.7,  the commutator  $[T_{f_1},T_{f_2}]$  is in the trace class, 
where  $f_1$ and  $f_2$   were defined by (4.1).  This allows us to 
apply the above theory to the pair  $A =  T_{f_1}$  and  $B = T_{f_2}$.  
Theorem 4.1 says that the essential spectrum of  $A + iB = T_F$  
is contained in   $\partial S$.  It follows from this fact that 
$$
\text{index}(T_F - \lambda )   = 0  \quad  \text{for every}  \ \  \lambda \in  {\bold C}\backslash S.
$$
\noindent
Therefore for this pair  $A =  T_{f_1}$  and  $B = T_{f_2}$,   we have   $g_{A,B} = n\chi _S$,  where  
$\chi _S$  is the characteristic function of the square  $S$  and  
$$
n =  \text{index}(T_F - \lambda )    \quad  \text{for each}  \ \  \lambda \in  S\backslash \partial S.
$$
\noindent
Applying (6.1) in the case  $p(x,y) = x$  and  $q(x,y) = y$,  we obtain
$$
\text{tr}[T_{f_1},T_{f_2}]   =  {-1\over 2\pi i}\iint n\chi _S(x,y)dxdy = {-n\over 2\pi i}.
\tag 6.3
$$
\noindent
The above two identities imply   
$$
\text{tr}[T_{f_1},T_{f_2}]  =  {-1\over 2\pi i}\text{index}(T_F - \lambda ) \quad  \text{for every}  \ \  \lambda \in  S\backslash \partial S.
\tag 6.4
$$
\noindent
This identifies the quantized Hall conductance with a Fredholm index in the case where the Fermi energy is strictly between 
the Landau levels  $\ell = 0$  and  $\ell = 1$.
\medskip
By Proposition 5.1, we have  $\text{tr}[T_{f_1},T_{f_2}] = (2\pi i)^{-1}$.  Compare this with (6.3),  
we find  that  $n = -1$,  i.e.,  
$$
\text{index}(T_F - \lambda ) = -1  \quad  \text{for every}  \ \  \lambda \in  S\backslash \partial S.
$$ 
In other words, for the pair  $A =  T_{f_1}$  and  $B = T_{f_2}$  we have   $g_{A,B} = -\chi _S$.  
Therefore, by (6.1),  we have
$$
\text{tr}[p(T_{f_1},T_{f_2}),q(T_{f_1},T_{f_2})]  =  {1\over 2\pi i}\iint _S\{p,q\}(x,y)dxdy
$$
\noindent
for all polynomials  $p, q \in {\bold C}[x,y]$.
\medskip
The fact that  $n \neq 0$  implies that no point of  $\partial S$  can be missing from the essential spectrum of   $T_F$.  This 
strengthens the case  $j = 0$  in Theorem 4.1 to the following extent:
\medskip
\noindent
{\bf Theorem  6.1.}  {\it On the Fock space}  ${\Cal F}^2$,  {\it the essential spectrum of the Toeplitz operator}  $T_F$   
{\it equals}  $\partial S$,  {\it the boundary of the square}  $S$. 
\medskip
With the Carey-Pincus theory in hand, we are now ready to generalize both Proposition 5.1 and formula (6.4) to the 
setting of the higher Fock spaces  ${\Cal F}_j$,  $j \geq 1$.

\bigskip
\centerline{\bf 7.  Trace calculation in the higher Fock spaces}
\medskip
First of all,  Proposition 3.7 implies that for every  $j \geq 0$,  the commutator 
$[T_{f_1,j},T_{f_2,j}]$  is in the trace class.  Our task in this section is to compute 
$\text{tr}[T_{f_1,j},T_{f_2,j}]$.  Given Proposition 5.1,  one obviously expects
$$
\text{tr}[T_{f_1,j},T_{f_2,j}]  =  {1\over 2\pi i}   \quad  \text{for every}  \ \  j \geq 0.
\tag 7.1
$$ 
\noindent
We will show that this is indeed true.  
\medskip
Because of (2.6),  for each  $j \geq 0$  we will write   $(j!)^{-1/2}C^jP = V_j$.  Thus  $V_j^\ast =  (j!)^{-1/2}PA^j$.  
Recall from (2.7)  that each  $V_j$  is a partial isometry.
\medskip
\noindent
{\bf Lemma 7.1.}   {\it Given any}  $j \geq 1$,  {\it there exist coefficients}    $c_1^{(j)}, \dots ,c_j^{(j)}$  
{\it such that if}    $f \in C^\infty ({\bold C})$   {\it and if}  $f$  {\it and} 
$\partial \bar \partial f, \dots ,\partial ^j\bar \partial ^jf$  {\it are all bounded on}  ${\bold C}$,  {\it then}
$$
V_j^\ast T_{f,j}V_j =  T_f  +  \sum _{\nu =1}^jc_\nu ^{(j)}T_{\partial ^\nu \bar \partial ^\nu f}.
$$
\medskip
\noindent
{\it Proof}.      First of all, since  $V_j = P_jV_j$   and  $V_j^\ast = V_j^\ast P_j$,  we have
$$
V_j^\ast T_{f,j}V_j  =  V_j^\ast P_jM_fP_jV_j =  V_j^\ast M_fV_j.
$$ 
\noindent
It is elementary that
$$
\align
A^jM_f  &=  M_fA^j + [A^j,M_f]   \\
&=  M_fA^j + M_{\bar \partial f}A^{j-1} + AM_{\bar \partial f}A^{j-2} + \cdots + A^{j-1}M_{\bar \partial f}  \\
&=  M_fA^j  + jM_{\bar \partial f}A^{j-1} + [A,M_{\bar \partial f}]A^{j-2} + \cdots + [A^{j-1},M_{\bar \partial f}]  \\
&= \cdots \cdots  \\
&= M_fA^j  +  \sum _{\nu =1}^jb_\nu ^{(j)}M_{\bar \partial ^\nu f}A^{j-\nu }. 
\endalign
$$
\noindent
By the commutation relation  $[A,C] =1$  and the fact  $APu = 0$  for every  $u \in {\bold C}[z,\bar z]$,  we have 
$A^{j-\nu }C^jP = (j!/\nu !)C^\nu P$,  $1 \leq \nu \leq j$.  Hence  
$$
\align
V_j^\ast M_fV_j &= {1\over j!}PA^jM_fC^jP   
= {1\over j!}PM_fA^jC^jP  +  \sum _{\nu =1}^j{b_\nu ^{(j)}\over j!}PM_{\bar \partial ^\nu f}A^{j-\nu }C^jP  \\
&=  T_f  +  \sum _{\nu =1}^jc_\nu ^{(j)}PM_{\bar \partial ^\nu f}C^\nu P.
\tag 7.2
\endalign
$$ 
\noindent
For  $\varphi \in C^\infty ({\bold C})$  such that  $\varphi $  and  $\partial \varphi $  are bounded on  ${\bold C}$,
we have  $[M_\varphi ,C] = M_{\partial \varphi }$.   Also, note that  $PC = (AP)^\ast = 0$.  
From these facts we deduce   $PM_{\bar \partial ^\nu f}C^\nu = PM_{\partial ^\nu \bar \partial ^\nu f}$  for every  $\nu \geq 1$.  
Substituting this in (7.2),  we find that
$$
V_j^\ast M_fV_j  =   T_f  +  \sum _{\nu =1}^jc_\nu ^{(j)}PM_{\partial ^\nu \bar \partial ^\nu f}P
=  T_f  +  \sum _{\nu =1}^jc_\nu ^{(j)}T_{\partial ^\nu \bar \partial ^\nu f}
$$
\noindent
as promised.    $\square $
\medskip
\noindent
{\bf Lemma 7.2.}    {\it Let}  $u$  {\it and}  $v$  {\it be real-valued}, {\it bounded measurable functions on}  ${\bold R}$.  {\it Define}
$$
\varphi (\zeta )  =  u(\text{Re}(\zeta ))   \quad  \text{and}  \quad  \psi (\zeta ) = v(\text{Re}(e^{-i\theta }\zeta )),
$$
\noindent
$\zeta \in {\bold C}$,  {\it where}  $\theta $  {\it is the same as in}  (4.1).  {\it Suppose that there is a}  $0 < \rho < \infty $  
{\it such that}  $u = 0$  {\it and}  $v = 0$  {\it on}  ${\bold R}\backslash (-\rho ,\rho )$.  {\it Then the commutators}
$$
[T_{f_1},T_\psi ],  \quad  [T_\varphi ,T_{f_2}]  \quad \text{\it and}  \quad  [T_\varphi ,T_\psi ]
$$
\noindent
{\it are in the trace class with zero trace}.
\medskip
\noindent
{\it Proof}.  We have  $v = v_+ - v_-$,  where   $v_+$  and  $v_-$   are non-negative, bounded 
measurable functions on  ${\bold R}$  which vanish on the set   ${\bold R}\backslash (-\rho ,\rho )$.  Thus there are 
$\xi _{+,1},\xi _{+,2}, \xi _{-,1}, \xi _{-,2} \in \Sigma _\rho $  and  coefficients  $c_+ $,  $c_-$    
such that  $v_+ = c_+(\xi _{+,1}- \xi _{+,2})$  and  $v_- = c_-(\xi _{-,1} - \xi _{-,2})$.   That is,
$$
v  =  c_+(\xi _{+,1}- \xi _{+,2}) -  c_-(\xi _{-,1} - \xi _{-,2}).  
$$
\noindent
Since  $\xi _{+,1},\xi _{+,2}, \xi _{-,1}, \xi _{-,2} \in \Sigma _\rho $,   
Proposition 3.7 implies that  $[T_{f_1},T_\psi ]$  is in the trace class, and Proposition 5.1 implies 
that the trace of  $[T_{f_1},T_\psi ]$  is zero.  Clearly,  $u$  admits a decomposition of the same kind.  
Therefore the other two commutators are also in the trace class with zero trace.   $\square $
\medskip
\noindent
{\bf Lemma 7.3.}    {\it Suppose that the functions}  $\eta $  {\it and}  $\xi $  {\it in}  (4.1)  {\it satisfy the condition}
$\eta , \xi \in \Sigma _a\cap C^\infty ({\bold R})$.    {\it Then} 
$$
\text{tr}[T_{f_1,j},T_{f_2,j}]  =  {1\over 2\pi i}
$$ 
\noindent
{\it for every}  $j \geq 1$.
\medskip
\noindent
{\it Proof}.    The condition  $\eta , \xi \in \Sigma _a\cap C^\infty ({\bold R})$  ensures that     
Lemma 7.1  is applicable to  $f_1$  and  $f_2$.   By that lemma,  for a given  $j \geq 1$,  we have 
$$
V_j^\ast [T_{f_1,j},T_{f_2,j}]V_j  =  [T_{f_1},T_{f_2}]  +  Z_1 + Z_2 + Z_3,
$$
\noindent
where
$$
\align
Z_1  &=   \sum _{\nu =1}^jc_\nu ^{(j)}[T_{f_1},T_{\partial ^\nu \bar \partial ^\nu f_2}],  \\
Z_2  &=   \sum _{\nu =1}^jc_\nu ^{(j)}[T_{\partial ^\nu \bar \partial ^\nu f_1},T_{f_2}]   \quad  \text{and}   \\
Z_3  &=   \sum _{\nu =1}^j\sum _{\nu '=1}^jc_\nu ^{(j)}c_{\nu '}^{(j)}
[T_{\partial ^\nu \bar \partial ^\nu f_1},T_{\partial ^{\nu '}\bar \partial ^{\nu '}f_2}].
\endalign
$$
\noindent
By Definition 3.5,  we have  $\eta = 0$  and  $\xi = 0$   on  $(-\infty ,-a)$  and  $\eta = 1$  and  $\xi = 1$   on  $(a,\infty )$.
Thus all the derivatives of  $\eta $  and  $\xi $  are supported on the interval  $[-a,a]$.  
Therefore each  $\partial ^\nu \bar \partial ^\nu f_1$  is a  $\varphi $  in Lemma 7.2,  and each  
 $\partial ^{\nu '}\bar \partial ^{\nu '}f_2$  is a  $\psi $  in Lemma 7.2.  
It follows from Lemma 7.2 that  $Z_1$,  $Z_2$  and  $Z_3$  are in the trace class with zero trace.  
Consequently,
$$
\align
\text{tr}[T_{f_1,j},T_{f_2,j}]  &=  \text{tr}([T_{f_1,j},T_{f_2,j}]P_j)  =  \text{tr}([T_{f_1,j},T_{f_2,j}]V_jV_j^\ast )   
=  \text{tr}(V_j^\ast [T_{f_1,j},T_{f_2,j}]V_j)  \\
&=  \text{tr}[T_{f_1},T_{f_2}]  +  \text{tr}(Z_1 + Z_2 + Z_3)  =   \text{tr}[T_{f_1},T_{f_2}].
\endalign
$$
Now an application of Proposition 5.1 completes the proof.    $\square $
\medskip
Since Proposition 3.7  tells us that the commutator   $[T_{f_1,j},T_{f_2,j}]$   is in the trace class,    
there is a Carey-Pincus principal function for the almost commuting pair   $T_{f_1,j}$,  $T_{f_2,j}$.
\medskip
\noindent
{\bf Lemma 7.4.}   {\it Suppose that the}  $\eta $  {\it and}  $\xi $  {\it in}  (4.1) {\it are arbitrary functions in}  $\Sigma _a$.   {\it Given a}   
$j \geq 1$,  {\it let}  $g_j$  {\it denote the Carey-Pincus principal function for the almost commuting self-adjoint operators}  
$T_{f_1,j}$   {\it and}  $T_{f_2,j}$.   {\it Then}
$$
g_j  =  -\chi _S.
$$ 
\medskip
\noindent
{\it Proof}.    First of all,  the existence of such a  $g_j$  is provided by (6.1).  
 Theorem 4.1 tells us that the essential spectrum of   $T_{F,j}$  is contained in  $\partial S$,  
whose two-dimensional Lebesgue measure is  $0$.  Thus by (6.2),  we have
$$
g_j  =  n_j\chi _S,
$$
\noindent
where  
$$
n_j  =  \text{index}(T_{F,j} - \lambda )    \quad \text{for every}  \ \  \lambda \in S\backslash \partial S.
$$
\noindent
The above holds true for an arbitrary pair of  $\eta ,\xi \in \Sigma _a$  in (4.1).
\medskip
Now take a pair of functions  $\tilde \eta ,\tilde \xi \in \Sigma _a\cap C^\infty ({\bold R})$.  Accordingly,  we define 
$$
\tilde f_1(\zeta )  =  \tilde \eta (\text{Re}(\zeta ))  \quad \text{and}  \quad  \tilde f_2(\zeta )  =  \tilde \xi (\text{Re}(e^{-i\theta }\zeta )),  
$$
\noindent
$\zeta \in {\bold C}$,  where the  $\theta $  is the same as in  (4.1).  Furthermore,  we define
$$
\tilde F  =  \tilde f_1 + i\tilde f_2.
$$
\noindent
By the preceding paragraph,  the almost commuting pair   $T_{\tilde f_1,j}$,  $T_{\tilde f_2,j}$   has a principal 
function  $\tilde g_j$,  and the principal function  $\tilde g_j$  has the form  $\tilde g_j = \tilde n_j\chi _S$,  where
$$
\tilde n_j  =  \text{index}(T_{\tilde F,j} - \lambda )    \quad \text{for every}  \ \  \lambda \in S\backslash \partial S.
$$
\noindent
Applying  Lemma 7.3  and identity  (6.1),  we have
$$
{1\over 2\pi i}  =  \text{tr}[T_{\tilde f_1,j},T_{\tilde f_2,j}] = {-\tilde n_j\over 2\pi i}\iint \chi _S(x,y)dxdy  =   {-\tilde n_j\over 2\pi i}.
$$
\noindent
From this we conclude that  $\tilde n_j = -1$.
\medskip
Next we show that  $n_j = \tilde n_j$,  and consequently  $n_j = -1$,  which proves the lemma.  
To prove that  $n_j = \tilde n_j$, we define
$$
\eta _t = t\eta + (1-t)\tilde \eta   \quad  \text{and}  \quad  \xi _t = t\xi + (1-t)\tilde \xi  ,  
$$
\noindent
$0 \leq t \leq 1$.  We then define,  for each  $0 \leq t \leq 1$,  the functions  
$$
f_{1,t}(\zeta )  =  \eta _t(\text{Re}(\zeta ))  \quad \text{and}  \quad   f_{2,t}(\zeta )  =   \xi _t(\text{Re}(e^{-i\theta }\zeta )), 
$$
\noindent
$\zeta \in {\bold C}$,   and  
$$
F_t  =  f_{1,t} + if_{2,t}.
$$
\noindent
Note that for every  $0 \leq t \leq 1$,  we have  $\eta _t, \xi _t \in \Sigma _a$.   Therefore by Theorem 4.1,  the essential spectrum of  
$T_{F_t,j}$  is contained in  $\partial S$,   $0 \leq t \leq 1$.  Moreover, the map  $t \mapsto T_{F_t,j}$  is obviously continuous 
with respect to the operator norm.  Therefore for each  $\lambda \in  S\backslash \partial S$,  the map
$$
t  \mapsto \text{index}(T_{F_t,j} - \lambda )
$$
\noindent
remains constant on the interval  $[0,1]$.  Since  $F_0 = \tilde F$  and  $F_1 = F$,  we have   $n_j = \tilde n_j$  
as promised.  This completes the proof.   $\square $
\medskip
\noindent
{\bf Proposition 7.5.}   {\it Suppose that the}  $\eta $  {\it and}  $\xi $  {\it in}  (4.1)  {\it are arbitrary functions in}  $\Sigma _a$.   
{\it Then for every}  $j \geq 0$  {\it we have} 
$$
\text{tr}[T_{f_1,j},T_{f_2,j}]  =  {1\over 2\pi i}.
$$  
\medskip
\noindent
{\it Proof}.   The case  $j = 0$  is covered by Proposition 5.1.   For  $j \geq 1$,  applying (6.1)  and Lemma 7.4,  we have
$$
\text{tr}[T_{f_1,j},T_{f_2,j}]  =  {-1\over 2\pi i}\iint g_j(x,y)dxdy =  {1\over 2\pi i}\iint \chi _S(x,y)dxdy  =  {1\over 2\pi i}.
$$
\noindent
$\square $
\medskip
\noindent
{\bf Remark.}  Using the method in this section, it is easy to verify that   $V_j^\ast T_{z/|z|, j}V_j$  is a compact 
perturbation of    $T_{z/|z|}$.    Since  $\text{index}(T_{z/|z|})$  is known to be  $-1$  (see [1] or Theorem 9.1 below),  we also 
have   $\text{index}(T_{z/|z|, j}) = -1$.   Thus Proposition 7.5 allows us to write
$$
\text{tr}[T_{f_1,j},T_{f_2,j}]  =  {-1\over 2\pi i}\text{index}(T_{z/|z|, j}),  
$$
\noindent
which generalizes (1.7) to each individual Landau level above the lowest one.   

\bigskip
\centerline{\bf 8. Hall conductance and Fredholm index}
\medskip
We will now put the results from the previous section together and identify the quantized  
Hall conductance, for the Landau levels below any $E$, with a Fredholm index.
\medskip
There is no obvious reason that  $\sigma _{\text{Hall}}$ should be additive with respect to orthogonal sum  
$P^{(\ell)}  =  \sum_{j=0}^\ell P_j$.  So the fact that it is additive, is not a trivial matter:
\medskip
\noindent
{\bf Theorem 8.1.}  {\it For each}   $\ell \geq 0$,  {\it the commutator}   $[T_{f_1}^{(\ell )},T_{f_2}^{(\ell )}]$  {\it is in the trace class with}
$$
\text{tr}[T_{f_1}^{(\ell )},T_{f_2}^{(\ell )}]  =  {\ell + 1\over 2\pi i}.
$$
\medskip
\noindent
{\it Proof}.   It follows from Proposition 3.7 that   $[T_{f_1}^{(\ell )},T_{f_2}^{(\ell )}]$  is in the trace class.
Define  
$$
\align
Z_0  &=  \sum \Sb 0\leq i,j,k\leq \ell  \\  i\neq j, j\neq k, k\neq i \endSb P_i[[M_{f_1},P_j],[M_{f_2},P_k]],  \\
Z_1  &=  \sum \Sb 0\leq i,j\leq \ell  \\  i\neq j\endSb P_i[[M_{f_1},P_i],[M_{f_2},P_j]],  \\
Z_2  &=  \sum \Sb 0\leq i,j\leq \ell  \\  i\neq j\endSb P_j[[M_{f_1},P_i],[M_{f_2},P_j]]     \quad  \text{and}    \\
Z_3  &=  \sum \Sb 0\leq i,j\leq \ell  \\  i\neq j\endSb P_i[[M_{f_1},P_j],[M_{f_2},P_j]].
\endalign
$$
\noindent
Since  $P^{(\ell )} = P_0 + \cdots + P_\ell $,  we have
$$
\align
[T_{f_1}^{(\ell )},T_{f_2}^{(\ell )}]  &=  P^{(\ell )}[[M_{f_1},P^{(\ell )}],[M_{f_2},P^{(\ell )}]]   \\
&=  \sum _{j=0}^\ell P_j[[M_{f_1},P_j],[M_{f_2},P_j]]   +  Z_0 + Z_1 + Z_2 + Z_3   \\ 
&=  \sum _{j=0}^\ell [T_{f_1,j},T_{f_2,j}]   +  Z_0 + Z_1 + Z_2 + Z_3, 
\endalign
$$
\noindent
By Proposition 3.7, the operators  $Z_0$,  $Z_1$,  $Z_2$,  $Z_3$  are in the trace class.  Applying
Proposition  7.5,  the proof will be complete once we show that   $\text{tr}(Z_0 + Z_1 + Z_2 + Z_3) = 0$.  
\medskip
Recall that the  $P_j$'s are orthogonal to each other.  Thus if  $i \neq j$,  $j \neq k$  and  $k \neq i$,  then
$$
P_i[[M_{f_1},P_j],[M_{f_2},P_k]]P_i  =  0.
$$
\noindent
Therefore  $\text{tr}(Z_0) = 0$.  
\medskip
For any pair of  $i \neq j$,  we have
$$
\align
P_i[[M_{f_1},P_j],[M_{f_2},P_j]]P_i  &=  P_i[M_{f_1},P_j][M_{f_2},P_j]P_i - P_i[M_{f_2},P_j][M_{f_1},P_j]P_i  \\
&=  -P_iM_{f_1}P_jM_{f_2}P_i  +  P_iM_{f_2}P_jM_{f_1}P_i  \\
&=  (P_iM_{f_2}P_j)(P_jM_{f_1}P_i)  -  (P_iM_{f_1}P_j)(P_jM_{f_2}P_i).  
\tag 8.1 
\endalign
$$
\noindent
Similarly,  if  $i \neq j$,  then
$$
P_j[[M_{f_1},P_i],[M_{f_2},P_i]]P_j  =  (P_jM_{f_2}P_i)(P_iM_{f_1}P_j)  -  (P_jM_{f_1}P_i)(P_iM_{f_2}P_j).   
\tag 8.2
$$
\noindent
By the famous Lidskii theorem, if  $X$  and  $Y$  are bounded operators such that both  $XY$  and  $YX$  are in the trace class,  
then  $\text{tr}(XY) = \text{tr}(YX)$.  Combining this fact with (8.1) and (8.2),  we reach the conclusion  $\text{tr}(Z_3) = 0$.
\medskip
For any pair of  $i \neq j$,  we have
$$
\align
P_i[[M_{f_1},P_i],[M_{f_2},P_j]]P_i  &=  P_i[M_{f_1},P_i][M_{f_2},P_j]P_i - P_i[M_{f_2},P_j][M_{f_1},P_i]P_i  \\
&=  P_iM_{f_1}(P_i -1)(-P_j)M_{f_2}P_i  -  P_iM_{f_2}P_jM_{f_1}P_i  \\
&=  (P_iM_{f_1}P_j)(P_jM_{f_2}P_i)  -  (P_iM_{f_2}P_j)(P_jM_{f_1}P_i). 
\tag 8.3   
\endalign
$$
\noindent
Similarly, when  $i \neq j$,  we have
$$
\align
P_j[[M_{f_1},P_i],[M_{f_2},P_j]]P_j  &=  P_j[M_{f_1},P_i][M_{f_2},P_j]P_j - P_j[M_{f_2},P_j][M_{f_1},P_i]P_j  \\
&=  P_jM_{f_1}P_iM_{f_2}P_j  -  P_jM_{f_2}(P_j-1)(-P_i)M_{f_1}P_j  \\
&=  (P_jM_{f_1}P_i)(P_iM_{f_2}P_j)  -  (P_jM_{f_2}P_i)(P_iM_{f_1}P_j).  
\tag 8.4   
\endalign
$$
\noindent
Combining (8.3) and (8.4) with the  Lidskii theorem,  we see that
$$
\text{tr}(P_i[[M_{f_1},P_i],[M_{f_2},P_j]]P_i + P_j[[M_{f_1},P_i],[M_{f_2},P_j]]P_j) = 0
$$
\noindent
whenever   $i \neq j$.  Hence  $\text{tr}(Z_1 + Z_2) = 0$.  This completes the proof.    $\square $
\medskip
Given an  $\ell \geq 0$,  let  $g^{(\ell )}$  denote the Carey-Pincus principal function for the almost commuting 
pair  $T_{f_1}^{(\ell )}$,  $T_{f_2}^{(\ell )}$.  It follows from  (6.2)  and Theorem 4.3 that
$$
g^{(\ell )}  =  n^{(\ell )}\chi _S,
$$
\noindent
where  
$$
n^{(\ell )}  =  \text{index}(T_F^{(\ell )} - \lambda )   \quad  \text{for every}  \ \  \lambda \in  S\backslash \partial S.
$$
\noindent
Applying  (6.1),  we have
$$
\text{tr}[T_{f_1}^{(\ell )},T_{f_2}^{(\ell )}]  =  {-n^{(\ell )}\over 2\pi i}\iint \chi_S(x,y)dxdy = {-n^{(\ell )}\over 2\pi i}.
$$
\noindent
Taking Theorem 8.1 and (2.9) into account, we can express the quantized Hall conductance  
$$
\sigma _{\text{Hall}}(P_{\leq E})  =   -i\text{tr}(P^{(\ell )}[[M_{f_1},P^{(\ell )}],[M_{f_2},P^{(\ell )}]])  
$$  
\noindent
for the case  $(2\ell +1)b < E < (2\ell + 3)b$,  $\ell \geq 0$, in the following two ways:
$$
\align
\sigma _{\text{Hall}}(P_{\leq E})  &=  {1\over 2\pi }\text{index}(T_{f_1+if_2}^{(\ell )} - \lambda ),   
\quad   \lambda \in  S\backslash \partial S;      \\
\sigma _{\text{Hall}}(P_{\leq E})  &=  - {\ell +1\over 2\pi }. 
\endalign
$$

\bigskip
\centerline{\bf 9.  Switch functions of a different kind}
\medskip
As was mentioned in the Introduction, we now consider the pair of functions
$$
\varphi _1(\zeta )  =  \text{Re}\bigg({\zeta \over |\zeta |}\bigg)   \quad  \text{and}  \quad
\varphi _2(\zeta )  =  \text{Im}\bigg({\zeta \over |\zeta |}\bigg),
$$
\noindent
$\zeta \in {\bold C}\backslash \{0\}$.   Furthermore, we define
$$
\Phi = \varphi _1 + i\varphi _2.
$$
\noindent
That is,  $\Phi (\zeta ) = \zeta /|\zeta |$  for  $\zeta \in {\bold C}\backslash \{0\}$.   
The Toeplitz operator  $T_\Phi $ is Fredholm, and its index is interpreted as the charge deficiency 
of the lowest Landau level when one magnetic flux quantum is introduced through the origin [1].
\medskip
Here is what we can prove in this situation:
\medskip
\noindent
{\bf Theorem 9.1.}   (1)  {\it The Toeplitz operator}  $T_\Phi $  {\it is a compact perturbation of the unilateral shift}.  

\noindent
(2)  {\it The commutator}   $[T_\Phi ^\ast ,T_\Phi ]$  {\it is in the trace class}.  {\it Consequently},  
{\it the commutator}   $[T_{\varphi _1},T_{\varphi _2}]$  {\it is in the trace class}.

\noindent
(3)  {\it We have}  $\text{tr}[T_\Phi ^\ast ,T_\Phi ] = 1$.  {\it In other words},  $\text{tr}[T_{\varphi _1},T_{\varphi _2}] = (2i)^{-1}$.
\medskip
\noindent
{\it Proof}.  We have the standard orthonormal basis  $\{e_k : k \geq 0\}$  for the Fock space  ${\Cal F}^2$,  where
$$
e_k(\zeta ) = (k!)^{-1/2}\zeta ^k.
$$
\noindent
Obviously,  we have  $\langle T_\Phi e_k,e_j\rangle = 0$  whenever  $j \neq k+1$.  On the other hand,  for each  $k \geq 0$,
$$
\align
\langle T_\Phi e_k,e_{k+1}\rangle &=  \langle \Phi e_k,e_{k+1}\rangle 
=  {1\over \pi \sqrt{k!(k+1)!}}\int {\zeta \over |\zeta |}\zeta ^k\overline{\zeta ^{k+1}}e^{-|\zeta |^2}dA(\zeta )    \\
&=  {2\over \sqrt{k!(k+1)!}}\int _0^\infty r^{2k+2}e^{-r^2}dr,
\tag 9.1
\endalign
$$
\noindent
which we will denote by   $a_k$.
\medskip
We claim that the following two statements hold true:

(a)  $a_{k+1} > a_k$  for every  $k \geq  0$.

(b)  $\lim _{k\rightarrow \infty }a_k = 1$.

\noindent
Postponing the proofs of (a), (b)  for a moment, we first deduce the conclusions of the theorem from these 
two statements.  
\medskip
By (9.1) and the fact that $\langle T_\Phi e_k,e_j\rangle = 0$  whenever  $j \neq k+1$,   we have  
$$
T_\Phi = \sum _{k=0}^\infty a_ke_{k+1}\otimes e_k.  
\tag 9.2
$$
\noindent
Thus (b) implies that  $T_\Phi $  is a compact perturbation of the unilateral shift
$$
V =  \sum _{k=0}^\infty e_{k+1}\otimes e_k,
$$
\noindent
proving (1).
\medskip
By (9.2),  we have  
$$
T_\Phi ^\ast T_\Phi = \sum _{k=0}^\infty a_k^2e_k\otimes e_k  \quad \text{whereas}  \quad  
T_\Phi T_\Phi ^\ast = \sum _{k=0}^\infty a_k^2e_{k+1}\otimes e_{k+1}.
$$
\noindent
Therefore
$$
[T_\Phi ^\ast ,T_\Phi ]  =   a_0^2e_0\otimes e_0 + \sum _{k=1}^\infty \{a_k^2 - a_{k-1}^2\}e_k\otimes e_k.
\tag 9.3
$$
\noindent
For    $m \geq 1$,  we have
$$
a_0^2 + \sum _{k=1}^m \{a_k^2 - a_{k-1}^2\}  =  a_m^2  \leq  1,
$$
\noindent
where the  $\leq $  follows from the fact that  $\|\Phi \|_\infty = 1$.  Combining this bound with the positivity  $a_k^2 - a_{k-1}^2 > 0$  
for every  $k \geq 1$, which is provided by (a),  we see that    $[T_\Phi ^\ast ,T_\Phi ]$  is in the trace class.
Once we know that  $[T_\Phi ^\ast ,T_\Phi ]$  is in the trace class,  we have
$$
\text{tr}[T_\Phi ^\ast ,T_\Phi ]  = a_0^2 + \lim _{m\rightarrow \infty }\sum _{k=1}^m \{a_k^2 - a_{k-1}^2\}  
=   \lim _{m\rightarrow \infty }a_m^2.
$$
\noindent
By (b), the right-hand side equals  $1$.  That is,  $\text{tr}[T_\Phi ^\ast ,T_\Phi ]  = 1$ as promised in (3).
\medskip
Let us now prove (a).  By (9.1),  for every  $k \geq 0$,
$$
\align
{a_{k+1}\over a_k}  &=  {\sqrt{k!(k+1)!}\over  \sqrt{(k+1)!(k+2)!}}\cdot 
{\int _0^\infty r^{2k+4}e^{-r^2}dr \over \int _0^\infty r^{2k+2}e^{-r^2}dr}  =  {k + (3/2)\over \sqrt{(k+1)(k+2)}}  \\
&=  \bigg({(k + (3/2))^2\over (k+1)(k+2)}\bigg)^{1/2}  =  \bigg({k^2 + 3k + (9/4)\over k^2 + 3k + 2}\bigg)^{1/2} > 1,
\endalign
$$
\noindent
where the second  $=$  involves an integration by parts.  This proves (a).
\medskip
To prove (b), first note that by (a),  the limit
$$
\lim _{k\rightarrow \infty }a_k    =  L 
$$
\noindent
exists.  Since  $\|\Phi \|_\infty = 1$,  we have  $a_k \leq 1$  for every  $k$.  Therefore  $L \leq 1$. 
The proof will be complete once we show that  $L \geq 1$.
\medskip
By the Cauchy-Schwarz inequality,  we have
$$
\bigg\{\int _0^\infty r^{2k+3}e^{-r^2}dr\bigg\}^2  \leq \int _0^\infty r^{2k+2}e^{-r^2}dr \int _0^\infty r^{2k+4}e^{-r^2}dr.
$$
\noindent
Combining this with (9.1),  we find that
$$
\align
a_ka_{k+1}  &=  {2\over \sqrt{k!(k+1)!}}\int _0^\infty r^{2k+2}e^{-r^2}dr{2\over \sqrt{(k+1)!(k+2)!}}\int _0^\infty r^{2k+4}e^{-r^2}dr  \\
&\geq {1\over (k+1)!\sqrt{k!(k+2)!}}\bigg\{2\int _0^\infty r^{2k+3}e^{-r^2}dr\bigg\}^2   
=  {\{(k+1)!\}^2\over (k+1)!\sqrt{k!(k+2)!}}  \\
&=  {k+1\over \sqrt{(k+1)(k+2)}}  =  \sqrt{k+1\over k+2}.
\endalign  
$$
\noindent
From this it is clear that    $L \geq 1$.  This completes the proof.   $\square $
\medskip
\noindent
{\bf Remark.}  By (a) and  (9.3),   the Toeplitz operator  $T_\Phi $ on  ${\Cal F}^2$   is {\it hyponormal}.
\medskip
By (1.4),  Theorem 9.1(3)  has the alternative form
$$
\text{tr}(P[[M_{\varphi _1},P],[M_{\varphi _2},P]])  =  {1\over 2i}.
\tag 9.4
$$
\noindent
Denote  $D = \{z \in {\bold C} :  |z| < 1\}$.  It is well known that for every  $\lambda \in D$,  $\text{index}(V-\lambda ) = -1$.  
Since  $T_\Phi $  is a compact perturbation of  $V$,  we have 
$$
\text{index}(T_\Phi - \lambda )  =  -1  \quad \text{for every}  \ \  \lambda \in D.
\tag 9.5
$$
\noindent
Thus we can rewrite (9.4) in the form
$$
\text{tr}(P[[M_{\varphi _1},P],[M_{\varphi _2},P]])  =  - {1\over 2i}\text{index}(T_\Phi - \lambda )  \quad \text{for every}  \ \  \lambda \in D.
\tag 9.6
$$
\noindent
This is the formula inspired by (1.7).
\medskip
Now consider the pair of operators  $A = T_{\varphi _1}$  and  $B = T_{\varphi _2}$.    By (6.2) and (9.5)  
and the fact  $\|T_\Phi \|  \leq 1$,  we have  
$$
g_{A,B}  =  - \chi _D
$$
\noindent
for the Carey-Pincus principal function for this pair.  By (6.1),  we have
$$
\text{tr}[p(T_{\varphi _1},T_{\varphi _2}),q(T_{\varphi _1},T_{\varphi _2})]  
=  {1\over 2\pi i}\iint _D\{p,q\}(x,y)dxdy  
$$
\noindent
for all  $p,q \in {\bold C}[x,y]$.  If we write the unilateral shift  $V$  in the form   $V = \tilde A + i\tilde B$,  then the above also gives us the identity
$$
\text{tr}[p(T_{\varphi _1},T_{\varphi _2}),q(T_{\varphi _1},T_{\varphi _2})]    
=  \text{tr}[p(\tilde A,\tilde B),q(\tilde A,\tilde B)]
$$
\noindent
for all  $p,q \in {\bold C}[x,y]$.
\medskip
Using the notation introduced in Section 2, we can rewrite (9.6) in the form
$$
\text{tr}(P^{(0)}[[M_{\varphi _1},P^{(0)}],[M_{\varphi _2},P^{(0)}]])  
=  - {1\over 2i}\text{index}(T_\Phi ^{(0)}- \lambda )  \quad \text{for every}  \ \  \lambda \in D.  
\tag 9.7
$$
\noindent
One is, of course, not completely satisfied with this.   
Given the results in Section 8, the obvious question is, what happens with this pair of  $\varphi _1$  and  $\varphi _2$  
at higher Landau levels?  The obvious guess is that (9.7) also holds for   $\ell \geq 1$.  But so far we have not been   
able to prove this.  Given (6.1) and (6.2), and given what we know about the Fredholm index,
all the mathematical difficulties can be reduced to a single problem:
\medskip
\noindent
{\bf Problem 9.2.}  For  $\ell \geq 1$,  does the commutator   $[T_{\varphi _1}^{(\ell )},T_{\varphi _2}^{(\ell )}]$  belong to 
the trace class?
\medskip
For the pair of switch functions  $f_1$  and  $f_2$  defined by (4.1),  we obtain the membership of   
$[T_{f_1}^{(\ell )},T_{f_2}^{(\ell )}]$  in trace class from Proposition 3.7.  In contrast, for  
$\varphi _1$  and  $\varphi _2$,  the individual terms  $[M_{\varphi _1},P][M_{\varphi _2},P]$  
and   $[M_{\varphi _2},P][M_{\varphi _1},P]$  are not in the trace class,  and Theorem 9.1(2)  depends on the 
cancellation between these terms.
If,  for  $\ell \geq 1$,  the commutator  
$[T_{\varphi _1}^{(\ell )},T_{\varphi _2}^{(\ell )}]$  is to be in the trace class, the right cancellation between the terms  
$$
[M_{\varphi _1},P^{(\ell )}][M_{\varphi _2},P^{(\ell )}]   \quad  \text{and}  \quad  [M_{\varphi _2},P^{(\ell )}][M_{\varphi _1},P^{(\ell )}] 
$$
\noindent
must take place to bring about this membership.  In other words, Problem 9.2 deals with a much more subtle situation.

\bigskip
\centerline{\bf 10.  A more classic example}
\medskip
Consider the pair of momentum, position operators   $D,M$   
that appear in every quantum mechanics textbook.   Namely, 
$$
D  =  -i{d\over dx},
$$
\noindent
which is a self-adjoint operator on a dense domain in   $L^2({\bold R})$,  and
$$
(Mh)(x) = xh(x),
$$
\noindent
which is also a self-adjoint operator on a dense domain in   $L^2({\bold R})$.  
For   $f, g \in C_c^\infty ({\bold R})$,  the operator   $f(M)g(D)$  is well known to be compact.  Thus  if  
$E$,  $F$  are bounded Borel sets in  ${\bold R}$,  then the operator
$$
\chi _E(M)\chi _F(D)
$$
\noindent
is compact.
\medskip
\noindent
{\bf Lemma 10.1.}  {\it Let}  $E$  {\it be a bounded Borel set in}  ${\bold R}$.  {\it Then the essential norm of}
$$
\chi _E(D) + \chi _E(M)
$$
\noindent
{\it is at most}  $1$.
\medskip
\noindent
{\it Proof}.  By the above,  the difference
$$
(\chi _E(D) + \chi _E(M)) - (\chi _E(D) + \chi _E(M))^2
$$
\noindent
is a compact operator.  Therefore the essential spectrum of  $\chi _E(D) + \chi (M)$  is contained in the 
two-point set  $\{0,1\}$.   $\square $
\medskip
Define the function
$$
\sigma (x)  =  {x\over \sqrt{1 + x^2}},    \quad  x \in {\bold R}.
$$
\noindent
It was shown in  [5]  that the commutator  $[\sigma (D),\sigma (M)]$  is in the trace class, and that
$$
\text{tr}[\sigma (D),\sigma (M)]  =  {2\over \pi i}.
\tag 10.1
$$
\noindent
See Example 2 in [5].   We will work out the Carey-Pincus theory for the pair  $\sigma (D)$, $\sigma (M)$.  
This shows that our techniques are capable of handling diverse examples.
\medskip
Let  $C_{\ast ,\ast }$  denote the collection of continuous functions   $f$  on  ${\bold R}$  such that both limits
$$
\lim _{x\rightarrow \infty }f(x)   \quad  \text{and}  \quad  \lim _{x\rightarrow -\infty }f(x) 
$$
\noindent
exist and are finite.  
\medskip
\noindent
{\bf Lemma 10.2.}  {\it For any}  $f,g \in C_{\ast ,\ast }$,  {\it the commutator}   $[f(D),g(M)]$  {\it is compact}.
\medskip
\noindent
{\it Proof}.  The maximal ideal space of   $C_{\ast ,\ast }$  is the obvious two-point compactification of  ${\bold R}$.  
The Gelfand transform of  $\sigma $  obviously separates points on this maximal ideal space. 
Thus by the Stone-Weierstrass approximation theorem,  
$\{p\circ \sigma : p \in {\bold C}[x]\}$  is dense in   $C_{\ast ,\ast }$  with respect to the norm  $\|\cdot \|_\infty $. 
Since   $[\sigma (D),\sigma (M)]$  is in the trace class,  therefore compact,   
the compactness of   $[f(D),g(M)]$,  $f,g \in C_{\ast ,\ast }$,   follows from from this density.   $\square $

\medskip
Let  $\Phi _0$  denote the collection of   continuous functions  $\varphi $  on  ${\bold R}$  satisfying the  following 
two conditions  

(1)   $-1 \leq \varphi \leq 1$  on   ${\bold R}$.

(2)  There is an  $R = R(\varphi ) \in (0,\infty )$  such that   $\varphi = 1$  on  $[R,\infty )$   and  
$\varphi = -1$  

on  $(-\infty ,-R]$.

\noindent
Define the square   $\Omega  = \{x+iy : x, y \in [-1,1]\}$.
\medskip
\noindent
{\bf Lemma 10.3.}  {\it If}  $\varphi ,\psi \in \Phi _0$,  {\it then the essential spectrum of the operator}
$$
T  =  \varphi (D) + i\psi (M)
$$
\noindent
{\it is contained in}  $\partial \Omega $.
\medskip
\noindent
{\it Proof}.   By Lemma 4.2,  it suffices to show that the essential spectrum of  $T$  
does not intersect the interior of  $\Omega $.
\medskip
Let  $\lambda = \alpha + i\beta \in \Omega \backslash \partial \Omega $,  i.e.,   $-1 < \alpha < 1$  and  $-1 < \beta < 1$.  
Then there are operators   $W$,  $X$,  $Y$,  $Z$   such that
$$
\align
\{1 - \alpha + i(\psi (M) - \beta )\}W  &= 1,   \\
\{-1-\alpha +i(\psi (M) - \beta )\}X  &=1,  \\
\{\varphi (D) - \alpha + i(1 - \beta )\}Y  &= 1   \quad \text{and}  \\
\{\varphi (D) - \alpha +i(-1 - \beta )\}Z  &= 1.
\endalign
$$
\noindent
By the membership  $\varphi , \psi \in \Phi _0$,  there is an   $R > 0$  such that   $\varphi = -1 = \psi $  on  $(-\infty ,-R]$  and  
$\varphi = 1 = \psi $  on  $[R,\infty )$.  Let  $\eta $  be a continuous function on  ${\bold R}$  satisfy the following conditions:  

(a)   $0 \leq \eta \leq 1$  on  ${\bold R}$.

(b)  $\eta = 0$  on   $(-\infty ,R+1]$.

(c)  $\eta = 1$  on  $[R+2,\infty )$.

\noindent
We further define the function  $\xi (x) = \eta (-x)$,  $x \in {\bold R}$.
\medskip
Since  $(\varphi - \alpha )\eta = (1 - \alpha )\eta $  on  ${\bold R}$,  we have
$$
\align
(T - \lambda )\eta(D)W  &=  (1-\alpha )\eta (D)W + i(\psi (M) - \beta )\eta (D)W  \\
&=   (1-\alpha )\eta (D)W + i\eta (D)(\psi (M) - \beta )W + K_1  \\
&=  \eta (D)\{(1-\alpha ) + i(\psi (M) - \beta )\}W + K_1 \\
&= \eta (D) + K_1,
\endalign
$$
\noindent
where   $K_1 = i[\psi (M),\eta (D)]W$,  which is a compact operator.  Similarly,  we have
$$
\align
(T - \lambda )\xi (D)X  &=  \xi (D) + K_2,  \\
(T - \lambda )\eta (M)Y  &=  \eta (M) + K_3  \quad  \text{and}  \\
(T - \lambda )\xi (M)Z  &=  \xi (M) + K_4,
\endalign
$$
\noindent
where   $K_2$,  $K_3$  and  $K_4$  are compact operators.  Thus if we define
$$
G  =  \eta(D)W + \xi (D)X +  \eta (M)Y + \xi (M)Z,
$$
\noindent
then
$$
(T - \lambda )G  =   \eta(D) + \xi (D) +  \eta (M) + \xi (M)  + K,
\tag 10.2
$$
\noindent
where  $K$  is a compact operator.  Define the function
$$
u  =  1 - \eta - \xi  
$$
\noindent
on  ${\bold R}$.   Then (10.2) becomes
$$
(T - \lambda )G  =  2 -  \{u(D) + u(M)\} + K.  
\tag 10.3
$$
\noindent
A chase of the definitions of  $\eta $, $\xi $  tells us that  $0 \leq u \leq 1$  on  ${\bold R}$  and that  
$u = 0$  on  $(-\infty ,- R-2]\cup [R+2,\infty )$.  Hence it follows from Lemma 10.1 that the essential norm of  $u(D) + u(M)$  
is at most  $1$.  Consequently,   $2 -  \{u(D) + u(M)\} + K$  is a Fredholm operator.  Therefore from (10.3) we conclude 
that  $T - \lambda $  has a right Fredholm inverse.  A similar argument shows that    $T - \lambda $  also has a left 
Fredholm inverse.  $\square $
\medskip
Let  $\Phi $  denote the collection of   continuous functions  $\varphi $  on  ${\bold R}$  satisfying the  following 
two conditions:  

(1)   $-1 \leq \varphi \leq 1$  on   ${\bold R}$.

(2)  $\lim _{x\rightarrow \infty }\varphi (x) = 1$  and   $\lim _{x\rightarrow -\infty }\varphi (x) = -1$.

\medskip
\noindent
{\bf Theorem 10.4.}  {\it If}  $\varphi ,\psi \in \Phi $,  {\it then the essential spectrum of the operator}
$$
T  =  \varphi (D) + i\psi (M)
$$
\noindent
{\it is contained in}  $\partial \Omega $.
\medskip
\noindent
{\it Proof}.   Again, by Lemma 4.2,  it suffices to show that the essential spectrum of  $T$  
does not intersect the interior of  $\Omega $.
\medskip
Let  $\lambda $  be a point in the interior of  $\Omega $.   For   $\varphi ,\psi \in \Phi $,  there are sequences  
$\{\varphi _k\} \subset \Phi _0$  and  $\{\psi _k\} \subset \Phi _0$  such that  $\lim _{k\rightarrow \infty }\|\varphi _k - \varphi \|_\infty = 0$
and  $\lim _{k\rightarrow \infty }\|\psi _k - \psi \|_\infty = 0$.  Define  
$$
T_k = \varphi _k(D)  + i\psi _k(M)
$$
\noindent
for each  $k$.  Then  $\lim _{k\rightarrow \infty }\|T - T_k\| = 0$.  By Lemma 10.3,  $T_k - \lambda $  is a Fredholm 
operator.  Let    $F_k$  be a Fredholm inverse of    $T_k - \lambda $.   Then
$$
T - \lambda =  T_k - \lambda + (T - T_k)  =  (T_k - \lambda )\{1 + F_k(T - T_k)\} + K_k,
\tag 10.4
$$
\noindent
where  $K_k$  is a compact operator.  If  $N$  is a normal operator whose spectrum is contained in  $\partial \Omega $,   
then  $\|(N - \lambda )^{-1}\| \leq 1/\text{dist}(\lambda ,\partial \Omega )$.   Lemma 10.2 tells us that  
$T_k$  is essentially normal.  Therefore by applying the GNS representation of the Calkin algebra and Lemma 10.3,    
we obtain the bound  $\|F_k\|_{\text{ess}} \leq 1/\text{dist}(\lambda ,\partial \Omega )$.  Thus by choosing a large  $k$,  
$\|F_k(T - T_k)\|_{\text{ess}}$  can be arbitrarily small.  Hence from (10.4) we see that   $T - \lambda $  has a Fredholm 
inverse.   $\square $
\medskip
We now apply the Carey-Pincus theory to the pair  $A =  \sigma (D)$  and  $B = \sigma (M)$.  
Theorem 10.4 says that the essential spectrum of  $\sigma (D) + i\sigma (M)$  
is contained in   $\partial \Omega $.  Hence   
$$
\text{index}(\sigma (D) + i\sigma (M) - \lambda )   = 0  \quad  \text{for every}  \ \  \lambda \in  {\bold C}\backslash \Omega .
$$
\noindent
Consequently, for the pair  $A =  \sigma (D)$  and  $B = \sigma (M)$,   we have   $g_{A,B} = n\chi _\Omega $,  where   
$$
n =  \text{index}(\sigma (D) + i\sigma (M) - \lambda )    \quad  \text{for every}  
\ \  \lambda \in  \Omega\backslash \partial \Omega .
$$
\noindent
Applying (6.1) in the case  $p(x,y) = x$  and  $q(x,y) = y$,  we obtain
$$
\text{tr}[\sigma (D),\sigma (M)]   =  {-1\over 2\pi i}\iint n\chi _\Omega (x,y)dxdy = {-2n\over \pi i}.
$$
\noindent
Comparing this with (10.1),  we find that  $n = -1$.   In other words,  
$$
\text{index}(\sigma (D) + i\sigma (M) - \lambda )   = -1   \quad  \text{for every}  \ \  
\lambda \in  \Omega \backslash \partial \Omega.
\tag 10.5
$$
\noindent
Thus  $-\chi _\Omega $  is the principal function for the pair  $A =  \sigma (D)$  and  $B = \sigma (M)$. 
Applying  (6.1)  in this situation,  we obtain the trace formula  
$$
\text{tr}[p(\sigma (D),\sigma (M)),q(\sigma (D),\sigma (M))]  =  {1\over 2\pi i}\iint _\Omega \{p,q\}(x,y)dxdy  
\tag 10.6
$$
\noindent
for all polynomials  $p, q \in {\bold C}[x,y]$.
\medskip
From (10.6) we see, for example, that  $\text{tr}([\sigma (D),\sigma (M)]\sigma (M)) = 0$.  Again, such a calculation 
would not be possible without the Carey-Pincus theory.
\medskip
\noindent
{\bf Theorem 10.5.}  {\it If}  $\varphi ,\psi \in \Phi $,  {\it then for every}  $\lambda \in  \Omega \backslash \partial \Omega $   
{\it we have}
$$
\text{index}(\varphi (D) + i\psi (M) - \lambda )   = -1.
$$
\medskip
\noindent
{\it Proof}.  Let  $\lambda \in  \Omega \backslash \partial \Omega $  be given.  For each  $0 \leq t \leq 1$,  define
$$
Y_t =  (t\varphi + (1-t)\sigma )(D)  +  i(t\psi + (1-t)\sigma )(M) - \lambda  .
$$
\noindent
By Theorem 10.4,  for every $0 \leq t \leq 1$,  the operator  $Y_t$  is Fredholm.   Obviously, the map  $t \mapsto Y_t$  
is continuous with respect to the operator norm.  Therefore the map  $t \mapsto \text{index}(Y_t)$  is a constant on  
$[0,1]$.  We have  $\text{index}(Y_0) = -1$  by (10.5).  Hence    $\text{index}(Y_1) = -1$.   $\square $
\medskip
Thus, by the above argument, whenever  $\varphi ,\psi \in \Phi $  are such that the commutator  
$[\varphi (D),\psi (M)]$  is in the trace class,    the principal function for the pair   $\varphi (D)$,  $\psi (M)$  
has no choice but be equal to  $-\chi _\Omega $,   which means 
$$
\text{tr}[p(\varphi (D),\psi (M)),q(\varphi (D),\psi (M))]  =  \text{tr}[p(\sigma (D),\sigma (M)),q(\sigma (D),\sigma (M))] .
$$   
\noindent
for all    $p, q \in {\bold C}[x,y]$.

\bigskip
\centerline{\bf Data availability}
\medskip
No data was used for the research described in the article.

\bigskip
\centerline{\bf References}
\medskip
\noindent
1.  Avron, J.E., Seiler, R., Simon, B.: Charge deficiency, charge transport and comparison of dimensions. 
Comm. Math. Phys. {\bf 159} 399-422 (1994)

\noindent
2.  Bellissard, J., van Elst, A., Schulz-Baldes, H.: The noncommutative geometry of the quantum Hall effect.
J. Math. Phys. {\bf 35} 5373-5451 (1994)

\noindent
3.  Carey, R.,   Pincus, J.:  An invariant for certain operator algebras.  Proc. Nat. Acad. Sci. U.S.A. {\bf 71} 1952-1956 (1974)

\noindent
4.  Carey, R.,   Pincus, J.:  Mosaics, principal functions, and mean motion in von Neumann algebras.  
Acta Math. {\bf 138}   153-218 (1977)

\noindent
5.  Elgart, A.,  Fraas, M.,  On Kitaev's determinant formula.  
St. Petersburg Math. J. {\bf 35}(1) 139-144 (2024)

\noindent
6.  Elgart, A., Graf, G.M., Schenker, J.H.: Equality of the bulk and edge Hall conductances in a mobility gap. 
Commun. Math. Phys. {\bf 259} 185-221 (2005)

\noindent
7.  Elgart, A., Schlein, B.: Adiabatic charge transport and the Kubo formula for Landau-type Hamiltonians. 
Comm. Pure Appl. Math. {\bf 57}(5) 590-615 (2004)

\noindent
8.  Helton, J.,   Howe, R.: Integral operators: commutators, traces, index and homology. 
Proceedings of a Conference on Operator Theory (Dalhousie Univ., Halifax, N.S., 1973), pp. 141-209. 
Lecture Notes in Math., {\bf 345}, Springer, Berlin, 1973

\noindent
9.  Ludewig, M., Thiang, G.C.: Quantization of conductance and the coarse cohomology of partitions. 
arXiv:2308.02819

\noindent
10.  Pincus, J.: Commutators and systems of singular integral equations I.  
Acta Math. {\bf 121}  219-249 (1968)

\noindent
11.  Thiang, G.C.: Index of Bargmann-Fock space and Landau levels. arXiv:2401.06660

\bigskip
\noindent
Guo Chuan Thiang

\noindent
Beijing International Center for Mathematical Research, Peking University,  Beijing 100871, China

\noindent
E-mail:  \url{guochuanthiang@bicmr.pku.edu.cn}

\bigskip
\noindent
Jingbo Xia

\noindent
College of Data Science, Jiaxing University,  Jiaxing 314001,  China

\noindent
and

\noindent
Department of Mathematics, State University of New York at Buffalo, Buffalo, NY 14260,  USA  

\noindent
E-mail: \url{jxia@acsu.buffalo.edu}

\end